\documentclass[12pt]{report}

\usepackage{tamuconfig}

\usepackage[T1]{fontenc}

\usepackage{graphicx}

\DeclareGraphicsExtensions{.png}

\graphicspath{ {./graphic/} }

\usepackage[pdftex,bookmarks,linktocpage,pdfpagelabels,plainpages=false,hyperfigures,linkcolor=black,citecolor=blue]{hyperref} 
\hypersetup{colorlinks=true}

\usepackage{amsmath}
\usepackage{amssymb}
\usepackage{enumerate}
\usepackage{amsmath}
\usepackage{tikz}
\usetikzlibrary{calc,trees,positioning,arrows,fit,shapes,calc}
\usepackage[compat=1.1.0]{tikz-feynman}
\usepackage[caption=false]{subfig}
\usepackage{feynmf}
\usepackage{slashed}
\usepackage{braket}
\usepackage{url}
\usepackage[sort&compress,numbers]{natbib}
\usepackage{tabularx}
\usepackage{multirow}

\usepackage{makecell}

\tikzstyle{decision} = [diamond, draw, fill=blue!20, 
    text width=4.5em, text badly centered, node distance=3cm, inner sep=0pt]
\tikzstyle{block} = [rectangle, draw, fill=blue!20, 
    text width=7em, text centered, rounded corners, minimum height=4em]
\tikzstyle{line} = [draw, -latex']
\tikzstyle{cloud} = [draw, ellipse,fill=red!20, node distance=3cm,
    minimum height=2em]

\usepackage{mathrsfs,amssymb}  
\usepackage{cancel}
\usepackage[normalem]{ulem}

\begin{document}

\renewcommand{\tamumanuscripttitle}{On Coherence in Bragg-Primakoff Axion Photoconversion}

\renewcommand{\tamupapertype}{Dissertation}

\renewcommand{\tamufullname}{Adrian Raphael Traquina Thompson}

\renewcommand{\tamudegree}{Doctorate of Science}
\renewcommand{\tamuchairone}{Bhaskar Dutta}

\renewcommand{\tamumemberone}{Louis Strigari}
\newcommand{\tamumembertwo}{Teruki Kamon}
\newcommand{\tamumemberthree}{Stephen A. Fulling}
\renewcommand{\tamudepthead}{Grigory Rogachev}

\renewcommand{\tamugradmonth}{August}
\renewcommand{\tamugradyear}{2023}
\renewcommand{\tamudepartment}{Physics}

%
%
%
%


\providecommand{\tabularnewline}{\\}

\begin{titlepage}
\begin{center}
\MakeUppercase{\tamumanuscripttitle}
\vspace{4em}

A \tamupapertype

by

\MakeUppercase{\tamufullname}

\vspace{4em}

\begin{singlespace}

Submitted to the Office of Graduate and Professional Studies of \\
Texas A\&M University \\

in partial fulfillment of the requirements for the degree of \\
\end{singlespace}

\MakeUppercase{\tamudegree}
\par\end{center}
\vspace{2em}
\begin{singlespace}
\begin{tabular}{ll}
 & \tabularnewline
& \cr
Chair of Committee, & \tamuchairone\tabularnewline
Committee Members, & \tamumemberone\tabularnewline
 & \tamumembertwo\tabularnewline
 & \tamumemberthree\tabularnewline
Head of Department, & \tamudepthead\tabularnewline

\end{tabular}
\end{singlespace}
\vspace{3em}

\begin{center}
\tamugradmonth \hspace{2pt} \tamugradyear

\vspace{3em}

Major Subject: \tamudepartment \par
\vspace{3em}
Copyright \tamugradyear \hspace{.5em}\tamufullname 
\par\end{center}
\end{titlepage}
\pagebreak{}

%
%
%
%

\chapter*{ABSTRACT}
\addcontentsline{toc}{chapter}{ABSTRACT} 

\pagestyle{plain} 
\pagenumbering{roman} 
\setcounter{page}{2}

Axions and axion-like pseudoscalar particles with dimension-5 couplings to photons exhibit coherent Bragg-Primakoff scattering with ordered crystals at keV energy scales. This provides for a natural detection technique in searches for axions produce in the Sun's interior. I will motivate the utility of dark matter direct detection experiments in searching for solar axions, emphasizing the role crystal-based detector technologies. I present an updated theoretical treatment of the Bragg-Primakoff photoconversion process for keV pseudoscalars, and address simultaneously the effects of absorption of final state photons in crystals on the loss of coherence, which can lead to large suppressive corrections to the event rate sensitivity for this detection technique. However, I also show that the Borrmann effect of anomalous absorption significantly lifts the suppression. This phenomenon is studied in Ge, NaI, and CsI crystal experiments and its impact on the the projected sensitivities of SuperCDMS, LEGEND, and SABRE to the solar axion parameter space. Lastly, I investigate the future reach of multi-ton scale crystal detectors and discuss strategies to maximize the discovery potential of experimental efforts in this vein.

\pagebreak{}

%
%
%
%

\chapter*{DEDICATION}
\addcontentsline{toc}{chapter}{DEDICATION}  

\begin{center}
\vspace*{\fill}
This body of work is the product of an arduous spirit that could not have been kindled, stoked, and tended without the love and support of, first and foremost, my mother Paula and father Kurt. And with the light of my friends, bright and fluorescent even in the darkest trench, I always had a familiar guide and tether to the things that were most important. For my advisor, Dr. Bhaskar Dutta, I have such a happy gratitude for the delicate balance of patience and ambition that was mentored. And without the kindness and enthusiasm of Dr. Teruki Kamon, whom I first worked under, I would not have started on this path. I want to deeply thank them, as well as my close collaborators Dr. James Dent, Dr. Doojin Kim, and Dr. Louis Strigari for engendering a creative environment where I was perfectly comfortable asking stupid questions and letting my mind run free without fear of judgement. Unfortunately that is a rare thing to find, sometimes, in our little community, and so I count myself very lucky for that and for all the people in my life. This work is dedicated to you all.
\vspace*{\fill}
\end{center}

\pagebreak{}

%
%
%
%

\chapter*{ACKNOWLEDGMENTS}
\addcontentsline{toc}{chapter}{ACKNOWLEDGMENTS}  

I am very grateful to Imran Alkhatib, Miriam Diamond, Amirata Sattari Javid, and John Sipe for the vigorous discussions and studies on the theoretical treatment of coherent Primakoff scattering in crystals and the comparison of numerical computations. I also graciously thank Tomohiro Yamaji for the insight on Laue-type diffraction, Timon Emken for the technical correspondence on the \texttt{DarkART} package, and Alexander Poddubny for the useful comments on Biagini's theory of anomalous absorption.

I also thank Nicole Bell, Kiwoon Choi, and Sebastian Hoof for discussions related to the work on inverse Primakoff scattering as it applied to dark matter direct detection experiments.

I would also like to thank the Mitchell Institute for Fundamental Physics and Astronomy for support. Portions of this research were conducted with the advanced computing resources provided by Texas A\&M High Performance Research Computing.

\pagebreak{}
%
%
%
%

\chapter*{CONTRIBUTORS AND FUNDING SOURCES}
\addcontentsline{toc}{chapter}{CONTRIBUTORS AND FUNDING SOURCES}  

\subsection*{Contributors}
This work was supported by a thesis committee consisting of Professor Bhaskar Dutta, Professor Louis Strigari, and Professor Teruki Kamon of the Department of Physics and Astronomy, and Professor Stephen A. Fulling of the Department of Mathematics.

Contributions to this work came from my close collaborators at Texas A\&M, Los Alamos National Laboratory, and abroad. The body of the thesis will include work from the following publications that I have authored:
\begin{enumerate}
\scriptsize
\item \textbf{J. B. Dent, B. Dutta, J. L. Newstead, \underline{A. Thompson}}, ``Inverse Primakoff Scattering as a Probe of Solar Axions at Liquid Xenon Direct Detection Experiments'' \textit{Phys. Rev. Lett.} \textbf{125} (2020) 13, 131805 \href{https://arxiv.org/pdf/2006.15118.pdf}{\tt arXiv:2006.15118}
\item \textbf{J. B. Dent, B. Dutta, \underline{A. Thompson}}, ``Bragg-Primakoff Axion Photoconversion in Crystal Detectors''. \textit{To appear.}
\end{enumerate}
All other work conducted for this thesis was completed by the student independently.
\subsection*{Funding Sources}
This work has been supported in part by DOE grant DE-SC0010813 and the Los Alamos National Laboratory LDRD program. I additionally thank the Mitchell Institute for Fundamental Physics and Astronomy for support.
\pagebreak{}

%
%
%
%

\phantomsection
\addcontentsline{toc}{chapter}{TABLE OF CONTENTS}  

\begin{singlespace}
\renewcommand\contentsname{\normalfont} {\centerline{TABLE OF CONTENTS}}

\setcounter{tocdepth}{4} 

\setlength{\cftaftertoctitleskip}{1em}
\renewcommand{\cftaftertoctitle}{%
\hfill{\normalfont {Page}\par}}

\tableofcontents

\end{singlespace}

\pagebreak{}


\phantomsection
\addcontentsline{toc}{chapter}{LIST OF FIGURES}  

\renewcommand{\cftloftitlefont}{\center\normalfont\MakeUppercase}

\setlength{\cftbeforeloftitleskip}{-12pt} 
\renewcommand{\cftafterloftitleskip}{12pt}

\renewcommand{\cftafterloftitle}{%
\\[4em]\mbox{}\hspace{2pt}FIGURE\hfill{\normalfont Page}\vskip\baselineskip}

\begingroup

\begin{center}
\begin{singlespace}
\setlength{\cftbeforechapskip}{0.4cm}
\setlength{\cftbeforesecskip}{0.30cm}
\setlength{\cftbeforesubsecskip}{0.30cm}
\setlength{\cftbeforefigskip}{0.4cm}
\setlength{\cftbeforetabskip}{0.4cm}



\listoffigures

\end{singlespace}
\end{center}

\pagebreak{}

%
\phantomsection
\addcontentsline{toc}{chapter}{LIST OF TABLES}  

\renewcommand{\cftlottitlefont}{\center\normalfont\MakeUppercase}

\setlength{\cftbeforelottitleskip}{-12pt} 

\renewcommand{\cftafterlottitleskip}{1pt}

\renewcommand{\cftafterlottitle}{%
\\[4em]\mbox{}\hspace{2pt}TABLE\hfill{\normalfont Page}\vskip\baselineskip}

\begin{center}
\begin{singlespace}

\setlength{\cftbeforechapskip}{0.4cm}
\setlength{\cftbeforesecskip}{0.30cm}
\setlength{\cftbeforesubsecskip}{0.30cm}
\setlength{\cftbeforefigskip}{0.4cm}
\setlength{\cftbeforetabskip}{0.4cm}

\listoftables 

\end{singlespace}
\end{center}
\endgroup
\pagebreak{}  

\pagestyle{plain} 
\pagenumbering{arabic} 
\setcounter{page}{1}

\chapter{INTRODUCTION}\label{ch:intro}

Axions and axion-like particles -- potentially long-lived pseudoscalars with weak couplings to the Standard Model (SM) that may have masses from the sub-eV to the GeV -- are central features in the landscape of solutions to the strong CP problem~\cite{Peccei:1977hh,Wilczek:1977pj}, and to the dark matter (DM) problem~\cite{Marsh:2015xka, Duffy:2009ig}, and otherwise appear ubiquitously in string theory~\cite{Svrcek:2006yi,Arvanitaki:2009fg,Cicoli:2012sz} - the so-called ``axiverse'' - and the ultraviolet (UV) spectra of many other puzzle-solving models. In this work I will dive into some of the mechanisms behind the aforementioned axion and ALP scenarios in order to motivate a broad target parameter space to search for ALPs.

While there has been a tremendous effort to probe high mass ALPs from the MeV to GeV scales, there is a deep chasm of parameter space for ALPs below 1 keV in mass that is neither constrained by laboratory probes nor astrophysics. In this region the ALP can act as both a dark matter candidate and a solution to the strong CP problem, among other roles in solutions to the current panorama of anomalies and puzzles. There are a number of future and ongoing helioscope experiments that use resonant cavities or large magnetic fields to search for DM or solar ALPs, but the reach of these experiments terminates around masses $\gtrsim 1$ eV. Dark matter direct detection experiments, on the other hand, offer broadband sensitivity in the axion mass and may be powerful enough at the tonne-scale to search for ALPs where the astrophysical constraints and future helioscopes and haloscopes lose their sensitivity. In particular, Bragg-Primakoff scattering -- a form of ALP-to-photon conversion that is coherent at the level of an ordered atomic lattice as well as coherent at the level of the atomic charge distribution -- can greatly enhance the event rates for ALPs with keV energies. This makes for a natural detection technique in the search for solar ALPs, and by leveraging crystal detectors we can get the reach we need from direct detection experiments at smaller detector scales. In this work I will thoroughly construct the methodology for this detection technique, building off the existing literature but repairing several stark inconsistencies along the way, in the end showing the importance and power of crystallographic technologies to search for axion-like particles with the aforementioned parameter space as a clear goal.

In Chapter~\ref{ch:theory} I discuss the theoretical motivations to search for axion-like particles, and in Chapter~\ref{ch:amplitudes} I will mark out the relevant amplitudes, cross sections, and form factors relevant for Primakoff scattering, the key phenomenological probe to look for axion-like particles coupling to photons. In Chapter~\ref{ch:crystals} I give a detailed treatment of Bragg-Primakoff conversion in perfect crystals, working through the subtleties and salient phenomenological impact of absorption effects (and the Borrmann effect) on the event rate for Bragg-Primakoff conversion. In Chapter~\ref{ch:solar} I apply the results of the previous two chapters to searches of axion-like particles produced in the Sun, and to other potential discovery strategies to test QCD axion model parameter space. Finally, in Chapter~\ref{ch:conclusion} I conclude and remark on future directions of work.
%
%
%
%



\chapter{\uppercase{Theoretical Foundations for Motivating the Existence of Axion-like Particles}}\label{ch:theory}

\section{Axionic Solutions to the Smallness of Strong $CP$}
We first begin with pseudoscalar particles that solve the strong CP problem, which I will label simply as ``axions'' or ``QCD axion'' and I will call pseudoscalars which do not necessarily solve strong CP, and span a much broader parameter space of SM couplings, ``axion-like particles'' or ALPs. The Peccei-Quinn mechanism~\cite{Peccei:1977hh,Wilczek:1977pj,Weinberg:1977ma,Preskill:1982cy,Abbott:1982af,Dine:1982ah,Duffy:2009ig,Marsh:2015xka,Battaglieri:2017aum}, whereby a global $U(1)_{PQ}$ symmetry is spontaneously broken, offers a dynamic mechanism to minimize the QCD vacuum to a CP-conserving state.

The way this happens can be understood by looking at the properties of the QCD vacuum. The QCD vacuum has two angles: $\theta$, the angle that appears in front of the $G\Tilde{G}$ topological term, and $\theta_q = \arg\det M_q$, the phase of the quark mass matrix. These combine as a physical parameter $\Bar{\theta} = \theta + \theta_q$ which has the consequence of violating $CP$;
\begin{equation}
    \mathcal{L} \supset \frac{g^2}{32 \pi^2} \bar{\theta} G \Tilde{G}
\end{equation}
Here I have used the abbreviated notation $G \Tilde{G} = \epsilon_{\alpha\beta\mu\nu} G^{\alpha\beta}_a G^{\mu\nu}_a$. The Vafa-Witten theorem states that if one considers two different configurations of the QCD vacuum, one where $\theta = 0$ and one where $\theta \geq 0$, the one set to zero will always have lower energy density. In a universe where $\theta$ is a fixed parameter, nothing would protect the theory from having $\theta \geq 0$ and violating CP; we would just have to throw up our hands and accept it. However, if $\Bar{\theta}$ were promoted to a dynamical field, then it would have the ability to roll down to the bottom of its potential at $\theta = 0$ where the Vafa-Witten theorem~\cite{PhysRevLett.53.535} tells us that is the energetically favored configuration. This is precisely what the axion field does for us, giving a mechanical explanation for why the observed $CP$ violation is so small.

\begin{figure}[h]
    \centering
\begin{tikzpicture}[node distance = 4cm, auto]
    \node (ksvz) {\underline{KSVZ-type}};
    \node [right of=ksvz, node distance=8cm] (dfsz) {\underline{DFSZ-type}};
    \node [below of=ksvz, node distance=1cm] (ksvz1) {Heavy Quarks + Singlet: $Q_L, Q_R, \Phi$};
    \node [below of=dfsz, node distance=1cm] (dfsz1) {2HDM + Singlet: $H_u, H_d, \Phi$};
    \node [below of=ksvz1, node distance=2cm] (ksvz2) {$\mathcal{L}\supset -m_Q \bar{Q}_L Q_R e^{-ia/v_a}$};
    \node [below of=ksvz2, node distance=2cm] (ksvz3) {$\mathcal{L}\supset a G\Tilde{G}, \, a F\Tilde{F}$ via $Q$ loops};
    \begin{feynman}
         \vertex [below=1cm of ksvz3] (o1);
         \vertex [left=0.5cm of o1] (i);
         \vertex [above right=0.5cm of o1] (f1);
         \vertex [below right=0.5cm of o1] (f2);
         \vertex [right=0.5cm of f1] (j1);
         \vertex [right=0.5cm of f2] (j2);

         \diagram* {
           (i) -- [scalar] (o1),
           (o1) -- (f1),
           (o1) -- (f2),
           (f1) -- (f2),
           (f1) -- [boson] (j1),
           (f2) -- [boson] (j2),
         };
    \end{feynman};

    \node [below of=dfsz1, node distance=2cm] (dfsz2) {$\mathcal{L}_{f=e,u,d} \supset -m_f \bar{f}_L f_R e^{-i\chi a/v_a}$};
    \node [below of=dfsz2, node distance=2cm] (dfsz3) {$\delta\mathcal{L}_{f=e,u,d} = \frac{c_{af}}{f_a} (\partial_\mu a) \bar{f}\gamma^\mu \gamma_5 f$};
    \node [below of=dfsz3, node distance=0.6cm] (dfsz4) {$\mathcal{L}\supset a G\Tilde{G}, \, a F\Tilde{F}$ via $e, u, d$ loops};
    \begin{feynman}
         \vertex [below left=1.2cm of dfsz4] (oo1);
         \vertex [left=0.5cm of oo1] (ii);
         \vertex [above right=0.5cm of oo1] (ff1);
         \vertex [below right=0.5cm of oo1] (ff2);
         \vertex [right=0.5cm of ff1] (jj1);
         \vertex [right=0.5cm of ff2] (jj2);

         \diagram* {
           (ii) -- [scalar] (oo1),
           (oo1) -- (ff1),
           (oo1) -- (ff2),
           (ff1) -- (ff2),
           (ff1) -- [boson] (jj1),
           (ff2) -- [boson] (jj2),
         };
    \end{feynman};
    \begin{feynman}
         \vertex [below right=1.2cm of dfsz4] (h1);
         \vertex [left=0.5cm of h1] (x1);
         \vertex [above right=0.7cm of h1] (g1);
         \vertex [below right=0.7cm of h1] (g2);

         \diagram* {
           (x1) -- [scalar] (h1),
           (h1) -- [fermion] (g1),
           (h1) -- [fermion] (g2),
         };
    \end{feynman};
    
    \path [line] (dfsz1) -- node {PQSB} (dfsz2);
    \path [line] (dfsz2) -- node {$f\to fe^{-i\chi\gamma^5 a/2v_a}$} (dfsz3);
    \path [line] (ksvz1) -- node {PQSB} (ksvz2);
    \path [line] (ksvz2) -- node {$Q\to Qe^{-i\gamma^5 a/2v_a}$} (ksvz3);
\end{tikzpicture}
\caption{The KSVZ and DFSZ type QCD axion model mechanisms summarized graphically.}
\label{fig:qcd_axion_models}
\end{figure}
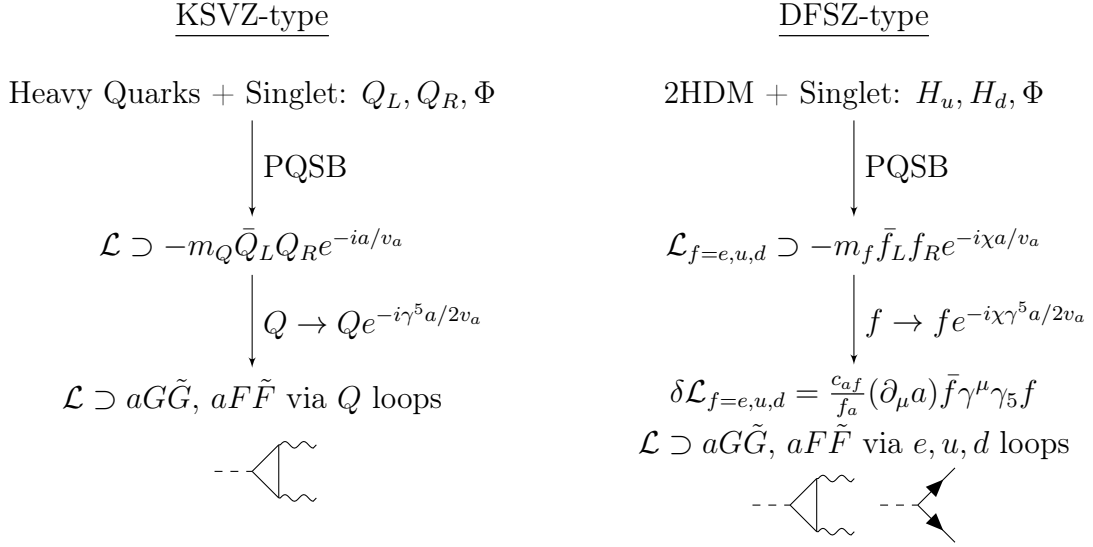

The next step in this line of thought is to construct a field $a$ that couples to $G\Tilde{G}$ and promotes $\bar{\theta}$ to a dynamical object. In Fig.~\ref{fig:qcd_axion_models} the common benchmark QCD axion models are shown schematically. Beginning with the Kim-Shifman-Vainshtein-Zakharov (KSVZ)~\cite{PhysRevLett.43.103, SHIFMAN1980493} type models on the left, which feature a set of heavy color-charged fermions $Q_L$ and $Q_R$ that are also charged under $U(1)_{PQ}$, derive couplings to gluons and photons through the operators $aG\Tilde{G}$ and $aF\Tilde{F}$ arising at loop level.

To see how this happens schematically, consider the PQ Yukawa terms like $\Phi \Bar{Q}_L Q_R$, which in the physical basis after $U(1)_{PQ}$ is broken become
\begin{equation}
    \Phi \Bar{Q}_L Q_R \to (v_a + \rho(x))e^{-i a / v_a}  \Bar{Q}_L Q_R,
\end{equation}
giving $Q$ a mass term and a coupling to the goldstone and radial modes of $\Phi$ in the broken phase. We can rotate away the goldstone angle $e^{-ia/v_a}$ by performing a field redefinition,
\begin{equation}
    Q_{L,R} \to Q_{L,R} e^{-i\gamma^5 a / 2 v_a}
\end{equation}
This brings a derivative coupling of the pseudoscalar axion field $a$ out of the $Q$ kinetic pieces after integrating by parts and throwing away the total derivative;
\begin{equation}
    i \bar{Q}_L \slashed{\partial} Q_R \to \frac{\partial_\mu a}{v_a} \bar{Q}_L \gamma^\mu \gamma^5 Q_R
\end{equation}

The Dine-Fischler-Srednicki-Zhitnitsky (DFSZ) type models~\cite{Zhitnitsky:1980tq,DINE1981199,Dine:1981rt,Dine:1982ah}, shown on the right, instead feature an extended Higgs sector which gives rise to tree-level fermion couplings after a chiral rotation of the fields, while couplings to gluons and photons are kept at loop level. Other variants exist, for example the sub-types DFSZ-I, DFSZ-II, IIa, IIb, etc.~\cite{Sun:2020iim} which categorize different schema for the Higgs sector structure. 

These different scenarios strongly motivate the existence of couplings to SM fermions and to the photon through the dimension-5 operator $a F \Tilde{F}$. The treatment of this parameter space can be expanded in a model-independent way by considering a simple effective field theory of ALP interactions;
\begin{equation}
    \mathcal{L} \supset \frac{1}{2}(\partial_\mu a )(\partial^\mu a) - m_a^2 a^2 - \dfrac{1}{4}g_{a\gamma} a F_{\mu\nu} \Tilde{F}^{\mu\nu} - \dfrac{1}{4}g_{ag} a \textrm{Tr}[G_{\mu\nu} \Tilde{G}^{\mu\nu}] - i  \sum_f g_{af} a\bar{f} \gamma^5 f 
\end{equation}
The QCD axion parameter space will then be a subset of the space covered by the axion mass and its couplings, as seen in Fig.~\ref{fig:gagamma_limits}.

The correlations between the QCD axion mass and its effective couplings are given below, taken from ref.~\cite{DiLuzio:2020wdo}. I will reiterate those correlations here for convenience of the reader. The relation between the Peccei-Quinn breaking scale $f_a$ and the axion mass is
\begin{equation}
    f_a =  \bigg(\frac{5.691\times 10^{6} \textrm{eV}}{m_a}\bigg) \textrm{GeV}
\end{equation}
To find the correlations between the axion mass and its effective couplings to photons in the KSVZ benchmark model is then given by Eq.~\ref{eq:ksvz_gagamma};
\begin{align}
\label{eq:ksvz_gagamma}
    g_{a\gamma} &= \frac{m_a}{\textrm{GeV}} \bigg(0.203 \frac{E}{N} - 0.39\bigg)
\end{align}
I will then take a region of model parameter space defined by considering anomaly number ratios from $E/N = 44/3$ to $E/N = 2$. This defines a band in the $(m_a, g_{a\gamma}$ parameter space of ALP couplings and masses.

For the DFSZ benchmark model, for which couplings to electrons would be dominant relative to the photon couplings, I take
\begin{equation}
\label{eq:dfsz_gae}
    g_{ae} = \dfrac{m_e C_{ae}(m_a, \tan\beta)}{f_a}
\end{equation}
where the coefficient $C_{ae}$ is dependent on the rotation angle $\beta$ for the vacuum expectation values of the extended Higgs sector in DFSZI and DFSZII models;
\begin{align}
    \textrm{DFSZ(I):\, \, }C_{ae} &= -\frac{1}{3} \sin^2\beta + \frac{3\alpha^2}{4 \pi^2} \bigg[\frac{E}{N}  \log(f_a/m_a) - 1.92 \log(1/m_e)\bigg],  \, \, \, \,   \frac{E}{N} = 8/3 \\
    \textrm{DFSZ(II):\, \,  } C_{ae} &= \frac{1}{3} \sin^2\beta + \frac{3\alpha^2}{4 \pi^2} \bigg[\frac{E}{N}  \log(f_a/m_a) - 1.92 \log(1/m_e)\bigg], \, \, \, \,  \frac{E}{N} = 2/3
\end{align}
Here I allow $\tan\beta$ values between 0.25 and 120, allowing for a wide range of coupling values for a particular axion mass~\cite{Giannotti:2017hny}.

As a brief aside on couplings to fermions: one can either have the Yukawa, or manifestly-pseudoscalar operator;
\begin{equation}
    \mathcal{L}_{Yuk} \supset i g_{ae} a \bar{\psi} \gamma^5 \psi
    \label{eq:yuk_op}
\end{equation}
or the gradient/derivative type operator associated with pNGBs with a shift symmetry;
\begin{equation}
    \mathcal{L}_{Grad} \supset  \frac{\partial_\mu a}{f_a} \bar{\psi} \gamma^\mu \gamma^5\psi
    \label{eq:grad_op}
\end{equation}
These two operators can be transformed into each other since they are related by a total derivative, but only if one assumes that the fermion $\psi$ satisfies its free equation of motion, i.e. $i \slashed{\partial} \psi = m \psi$. Therefore, in general the two operators do not give equivalent amplitudes except when fermions can be approximated as on-shell.

While searches for QCD axions, that is, ALPs which solve the strong CP problem as part of a Peccei-Quinn mechanism, are well motivated, it is important to also consider alternative solutions. For example, left-right symmetric models have also been able to solve strong CP without needing an explicit axion. This only offers more motivation to search for them, since if a discovery is not made, and instead the viable parameter space for QCD axions is ruled out by experiment, then it will give us the invaluable guidance to look elsewhere for non-axionic solutions which we already know are possible.

Besides the scope of QCD axions mentioned here so far, there is a much broader landscape of axion-like particle solutions to modern physics puzzles. String \textit{axiverse} pseudoscalars arise from string theory compactification scenarios, which may number in the tens to hundreds from string theory compactification scenarios~\cite{Arvanitaki:2009fg,Cicoli:2012sz}. More generally, light pseudoscalars with weak couplings to the SM could arise as pNGBs from other broken symmetries such as $B-L$ or $T3R$, and may be introduced to address other important problems in the SM (e.g. Majorons, Familons). We should keep this in mind when motivating tests of ALP parameter space, having the understanding that the target space is often much larger and open ended than that of the traditional QCD models.

\section{Laboratory Probes}
\begin{figure}[h]
    \centering
    \includegraphics[width=1.0\textwidth]{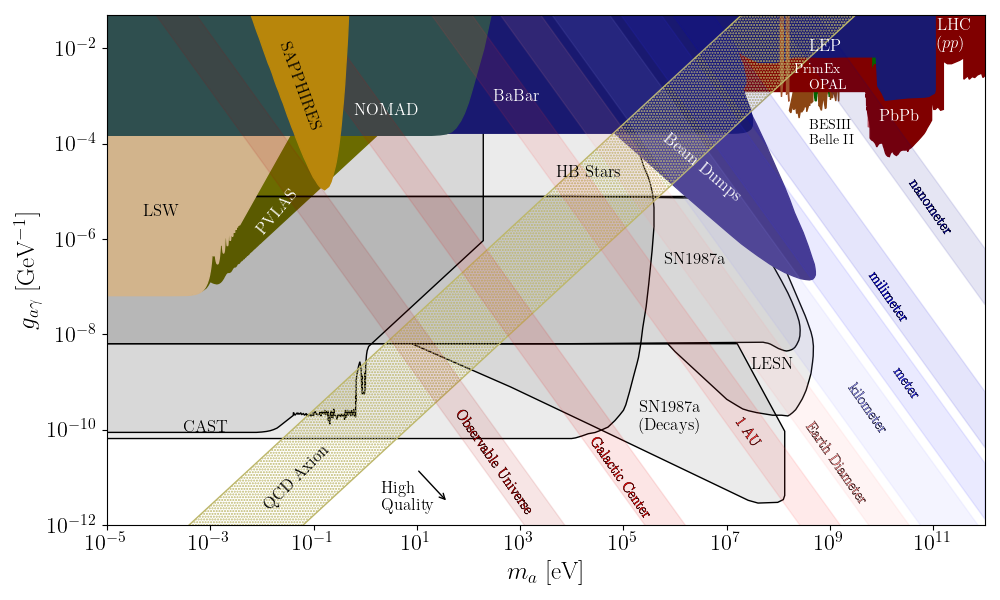}
    \caption{Existing ALP limits and constraints for ALPs coupling to photons through $a F \Tilde{F}$. Laboratory-derived constraints are shaded in solid colors while astrophysical constraints appear in grayscale. Bands of constant decay length are highlighted, whose thicknesses indicate a range of non-relativistic to relativistic ALPs. Traditional QCD axion model parameter space is shaded in yellow, corresponding to the KSVZ and DFSZ benchmark models. Additional constraints from cosmological considerations and assuming a dark matter ALP are not shown.}
    \label{fig:gagamma_limits}
\end{figure}
In Fig.~\ref{fig:gagamma_limits} I show many of the existing limits that have been set on the $g_{a\gamma} - m_a$ parameter space. Some of the most stringent constraints come from astrophysics, shown in gray, like those from the axion-induced stellar cooling of horizontal branch (HB) stars~\cite{Raffelt:1996wa}, supernova SN1987a energy loss~\cite{Lucente:2020whw,Caputo:2021rux,Caputo:2022mah,Payez:2014xsa,Hoof:2022xbe} and decaying axion signals~\cite{Jaeckel:2017tud,Hoof:2022xbe,Muller:2023vjm}, and low energy supernovae~\cite{Caputo:2022mah}. Pure laboratory constraints -- that is, those from searches for axions that would be produced and detected terrestrially -- are shown with solid colors~\cite{AxionLimits}. They include collider searches like those from LEP~\cite{Jaeckel:2015jla}, ATLAS~\cite{ATLAS:2020hii} and CMS from $pp$ collisions as well as Pb-Pb collisions~\cite{CMS:2018erd}, decaying particle lifetime measurements like those from PrimEx~\cite{PrimEx:2010fvg}, Belle II~\cite{Dolan:2017osp, Belle-II:2020jti}, BESIII~\cite{BESIII:2022rzz}. Beam dump experiments have also played a significant role in testing parameter space of axions which decay on the scale of the experimental baseline, namely E137~\cite{Dolan:2017osp}. Myself and collaborators exploited the intense electromagnetic and baryonic particle cascades inside beam dump targets and reactor cores could be used in searches that would be sensitive to high-mass ALPs in this parameter space~\cite{Dent:2019ueq,Brdar:2020dpr,Brdar:2022vum,CCM:2021jmk,Waites:2022tov}.

To keep a reference of scale, I also show bands of constant decay length to photon pairs, via the decay width
\begin{equation}
    \Gamma(a \to \gamma\gamma) = \dfrac{g_{a\gamma}^2 m_a^3}{64\pi}
    \label{eq:photon_width}
\end{equation}
The bands range from the size of the observable universe to nanometer scale, where the width of each band sweeps from non-relativistic ($\beta \sim 0.01$) to relativistic ($\beta > 0.9$) ALPs. The model parameter space for traditional QCD axions (KSVZ, DFSZ) is shown by the yellow band, however a much broader parameter space is motivated by non-traditional QCD axion models or by non-QCD ALP models. For example, many non-traditional QCD axion models are motivated to ensure that higher-order operators in the ultraviolet scales, which break $PQ$ symmetry, do not misalign the axion potential, spoiling the dynamical minimization of CP violation in the QCD vacuum. This has been known as the \textit{axion quality problem}~\cite{Kamionkowski:1992mf, Dine:1986bg, Georgi:1981pu, PhysRevD.46.539, Ghigna:1992iv}. Many such models have been realized which can shift the axion mass parametrically lighter~\cite{Elahi:2023vhu} or heavier~\cite{Gaillard:2018xgk, Kivel:2022emq, Hook:2019qoh}. These are just some examples, and of course the model parameter space of generic pNGBs or axiverse pseudoscalars is much larger. The richness of possibilities makes the physics targets for searches of ALPs also a broad eneavor.

While the frontier for axion searches in the high mass regime is busy with activity, the low mass regime is also very active but has its own challenges. \textit{Helioscope} experiments, like the International Axion Observatory (IAXO)~\cite{Armengaud:2019uso}, aim to probe sub-eV masses of ALPs associated with solar axions as the next generation successor to CAST~\cite{CAST:2017uph}. At the same time, a multitude of resonant cavity \textit{haloscopes} and other coherent electromagnetic-based technologies search for axions that might make up the local dark matter halo. These experiments cover a wide swath of parameter space for light axions below the eV scale. However, there is still a critical region of interest just at the eV scale which hosts several theoretical targets. Looking again at Fig.~\ref{fig:gagamma_limits}, we have (i) traditional QCD axion models, which are unconstrained from HB stars just below $m_a = 1$ eV, (ii) a number of anomalies in the cooling and growth rates of stars, which can be explained by ALPs with photon couplings just below $g_{a\gamma} \lesssim 10^{-11}$ GeV$^{-1}$~\cite{Giannotti:2015kwo,Hoof:2018ieb,Ayala:2014pea} (not pictured), and (iii) the lifetimes of ALPs in this same region of parameter space have lifetimes beyond that of the age of the universe, making them excellent dark matter candidates. It is this unique region of parameter space that is the prime target of this thesis, and in the following chapters, I will motivate the phenomenology of crystal detector technology that leverages coherence to be sensitive to ALP detection in this context.

%
%
%
%

\chapter{\uppercase{Electromagnetic-Axion Interactions}}\label{ch:amplitudes}
In this section I will review the production and detection processes that are relevant to laboratory searches, in particular the Primakoff and inverse Primakoff photoconversion processes relevant for axions coupled to photons, in addition to a discussion on other processes relevant for the production of axions in the Sun.

The form of the vertex operator for the dimension-5 photon coupling is interesting, as the derivatives contained the electromagnetic field strength tensors give rise to momentum dependence in the Feynman rules. Here I provide a quick derivation. The operator concerning the $a\gamma \gamma$ vertex is
\begin{equation}
    \mathcal{L}_{a\gamma} = \dfrac{1}{4}g_{a\gamma} a F_{\mu\nu} \tilde{F}^{\mu\nu}
\end{equation}
where $ \tilde{F}^{\mu\nu} = \epsilon^{\mu\nu\alpha\beta}F_{\alpha\beta}$. Expanding out the operator and moving to momentum space for the vertex $a(k) \to \gamma(p) \gamma (q)$ yields
\begin{align}
    a F_{\mu\nu} \tilde{F}^{\mu\nu} &= \epsilon^{\mu\nu\alpha\beta} a(k) \bigg(\partial_\mu A_\nu(p) - \partial_\nu A_\mu (p) \bigg) \bigg( \partial_\alpha A_\beta(q) - \partial_\beta A_\alpha (q)\bigg) \nonumber \\
    &= - \epsilon^{\mu\nu\alpha\beta} a(k) \bigg(p_{\mu} A_\nu - p_{\nu} A_\mu \bigg) \bigg( q_{\alpha} A_\beta - q_{\beta} A_\alpha\bigg) \nonumber \\
    &= -a(k) \bigg( \epsilon^{\mu\nu\alpha\beta}(p_\mu A_\nu q_\alpha A_\beta - p_\mu A_\nu q_\beta A_\alpha) - \epsilon^{\mu\nu\alpha\beta}p_\nu A_\mu q_\alpha A_\beta + \epsilon^{\mu\nu\alpha\beta} p_\nu A_\mu q_\beta A_\alpha \bigg) \nonumber
\end{align}
Now relabel $\mu \Longleftrightarrow \nu$ and commute the Levi-Civita indices on the last two terms:
\begin{align}
    &= -2 \epsilon^{\mu\nu\alpha\beta} a(k) p_\mu A_\nu \bigg( q_\alpha A_\beta - q_\beta A_\alpha \bigg) \nonumber \\
    &= -4 \epsilon^{\mu\nu\alpha\beta} a(k) p_\mu A_\nu (p) q_\alpha A_\beta (q)
\end{align}
To get the vertex function from here, we take
$$\dfrac{(-i)^3 \delta^3}{\delta a(k) \delta A^\nu (p) \delta A^\beta (q)},$$
This gives us the final expression (and after multiplying back in the factor of $\frac{1}{4} g_{a\gamma}$), which is
\begin{equation}
    i g_{a\gamma} \epsilon^{\mu\nu\alpha\beta} q_\alpha p_\beta
\end{equation}
This is the vertex rule for the ALP-photon coupling, just as expected by analogy to the $\pi^0\gamma\gamma$ interaction. I will use this vertex rule to derive amplitudes for several key processes involving the photon coupling in the next few sections.

\begin{figure}[h]
    \centering
     \begin{tikzpicture}
       \begin{feynman}
         \vertex (o1);
         \vertex [left=1.4cm of o1] (i) {\(a\)};
         \vertex [above right=1.4cm of o1] (f1) {\(\gamma\)};
         \vertex [below right=1.4cm of o1] (f2) {\(\gamma\)};

         \diagram* {
           (i) -- [scalar] (o1),
           (o1) -- [boson] (f1),
           (o1) -- [boson] (f2),
         };
        \end{feynman}
       \end{tikzpicture}
    \caption{ALP decays through $a \to \gamma \gamma$ via the $g_{a\gamma}$ coupling.}
    \label{fig:gammagamma_decay}
\end{figure}
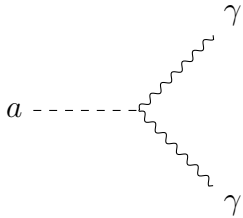

First we discuss ALP decays $a \to \gamma \gamma$ through the $g_{a\gamma}$ coupling (Fig.~\ref{fig:gammagamma_decay}).
The decay width is given by Eq.~\ref{eq:photon_width} which has a very strong cubic dependence on the ALP mass. This decay is kinematically accessible at all masses, but becomes only relevant for masses $m_a \gtrsim 10$ keV, otherwise the lifetime will be greater than about the age of the universe and becomes to rare to be observed easily.

\section{Atomic Primakoff Scattering}
The Primakoff scattering process has been made famous in the search for axions and ALPs. Originally named after $\pi^0 \to \gamma$ photoconversion process~\cite{Primakoff:1951iae,PhysRevLett.33.1400}, ALP Primakoff scattering is identical at the operator level and was investigated in the context of $a \to \gamma$ conversion through the coherent scattering with the strong atomic electric field of the nucleus and the electron cloud~\cite{Tsai:1986tx}. The coherent scattering with the entire atomic and nuclear charge density via a $t$-channel photon means that the cross section enjoys a proportionality to the square of the atomic number, $\propto Z^2$. But this is not the only form of coherence! Primakoff conversion may also sometimes refer to the same process taking place in the presence of a macroscopic electric or magnetic field, where in each case the coherency is still manifest but on different length scales~\cite{PhysRevLett.51.1415}. These make for the design principles behind resonant cavity haloscopes and helioscopes look for axions in the dark matter halo or those that might be free-streaming from the sun converting into electromagnetic signals on the Earth. In this thesis, though, I will contrast these search methods with a focus on atomic coherence, and later, Bragg or Laue coherence at the level of ordered crystals.
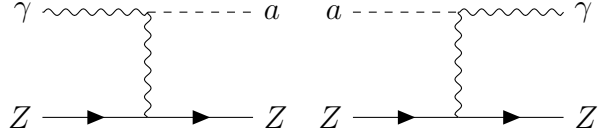
\begin{figure}
    \centering
    \begin{tikzpicture}
              \begin{feynman}
         \vertex (o1);
         \vertex [right=1.4cm of o1] (f1) {\(a\)};
         \vertex [left=1.4cm of o1] (i1){\(\gamma\)} ;
         \vertex [below=1.4cm of o1] (o2);
         \vertex [right=1.4cm of o2] (f2) {\(Z\)};
         \vertex [left=1.4cm of o2] (i2) {\(Z\)};

         \diagram* {
           (i1) -- [boson] (o1) -- [scalar] (f1),
           (o1) -- [boson] (o2),
           (i2) -- [fermion] (o2),
           (o2) -- [ fermion] (f2),
         };
        \end{feynman} 
       \end{tikzpicture}
       \begin{tikzpicture}
              \begin{feynman}
         \vertex (o1);
         \vertex [right=1.4cm of o1] (f1) {\(\gamma\)};
         \vertex [left=1.4cm of o1] (i1){\(a\)} ;
         \vertex [below=1.4cm of o1] (o2);
         \vertex [right=1.4cm of o2] (f2) {\(Z\)};
         \vertex [left=1.4cm of o2] (i2) {\(Z\)};

         \diagram* {
           (i1) -- [scalar] (o1) -- [boson] (f1),
           (o1) -- [boson] (o2),
           (i2) -- [fermion] (o2),
           (o2) -- [fermion] (f2),
         };
        \end{feynman}
       \end{tikzpicture}
    \caption{Primakoff scattering (left) and inverse-Primakoff scattering (right) from an atom of atomic number $Z$.}
    \label{fig:primakoff}
\end{figure}

The matrix element for Primakoff scattering with a free, heavy fermion $a(k) N(p) \to \gamma(k^\prime) N(p^\prime)$ with momentum transfer $q = k - k^\prime$, can be written down using the Feynman rules;
\begin{equation}
    \mathcal{M}_\textrm{free} = \Bar{u}(p^\prime) (-i e \gamma^\mu) u(p) \bigg(\dfrac{-ig_{\mu\nu}}{q^2} \bigg) (i g_{a\gamma} \epsilon^{\nu\rho\alpha\beta} q_\alpha k^\prime_\beta) \varepsilon^*_\rho (k^\prime)
\end{equation}
Squaring and evaluating the trace in terms of the usual Mandelstam variables $t = -q^2$ and $s = (k + p)^2$, and averaging over initial state spins, yields
\begin{equation}
\label{eq:primakoff_m2}
 \braket{|\mathcal{M}_\textrm{free}|^2} =  \frac{8\pi \alpha g_{a\gamma}^2}{t^2} \bigg[m_a^2 t \left(M^2+s\right)-t \left(\left(s-M^2\right)^2+s t\right)-M^2 m_a^4 -\frac{1}{2} t \left(t-m_a^2\right)^2\bigg]
\end{equation}
To apply this to a real atomic target, we can factorize the free matrix element and the atomic form factor separately to compute the interaction with the nuclear and electron cloud charge density in a straightforward way, as
\begin{equation}
    \braket{\mid \mathcal{M}\mid^2} = \braket{\mid \mathcal{M}\mid^2}_\textrm{free} \times F^2(q)
\end{equation}
For a hydrogenic potential, we use Tsai's parameterization~\cite{Tsai:1986tx} of the atomic form factor $F^2(t)$,
\begin{equation}
\label{eq:atomic_ff}
    F_A^2(q) = \dfrac{Z^2 a^4 q^4}{(1+a^2 q^2)^2}
\end{equation}
with $a=184.15 e^{-1/2} Z^{-1/3} / m_e$. For more context about how this form factor is constructed, I've added some notes in Appendix~\ref{app:coh_ff}.

In the forward limit, for an axion of momentum $k_a$, the inverse Primakoff cross section is given by~\cite{Avignone:1988bv,Creswick:1997pg,Avignone:1997th}
\begin{equation}
    \sigma(k_a) = \dfrac{Z^2 \alpha g_{a\gamma}^2}{2} \bigg(\dfrac{2 r_0^2 k_a^2 + 1}{4 r_0^2 k_a^2} \ln \Big(1+4r_0^2 k_a^2 \Big) - 1 \bigg)
    \label{eq:xs}
\end{equation}

\begin{figure}[h]
    \centering
    \includegraphics[width=0.6\textwidth]{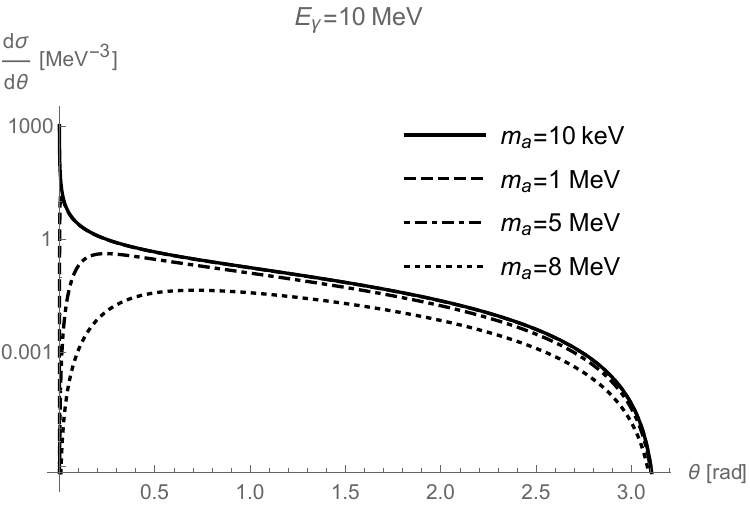}
    \includegraphics[width=0.6\textwidth]{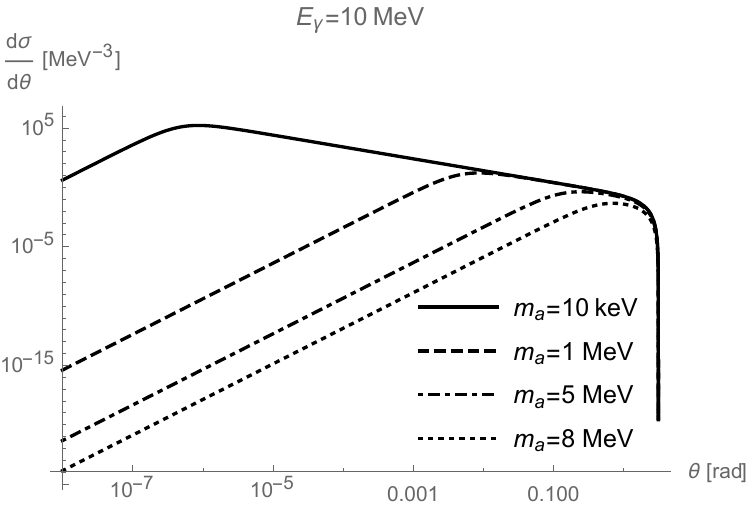}
    \caption{Primakoff scattering ($\gamma A \to a A$) differential cross section $d\sigma / d\theta$ as a function of the scattering angle $\theta$ with $g_{a\gamma}=1$ MeV$^{-1}$ and $Z=1$, on a log-linear graph (top) and log-log (bottom).}
    \label{fig:primakoff_diff_xs}
\end{figure}

Primakoff scattering is a very forward process; since the momentum transfer is very small, the typical scattering angles of the outgoing photon or ALP relative to the incoming momentum are typically small. This changes, however, as the ALP becomes non-relativistic, as shown in Fig.~\ref{fig:primakoff_diff_xs}.

%
%
%
%

\newcommand{\vb}[1]{\boldsymbol{#1}}
\newcommand{\polvec}{\vb\varepsilon}

\chapter{\uppercase{Axion-crystal Interactions}}\label{ch:crystals}

Axion-like particles in the keV to sub-eV mass range produced in the sun are well motivated~\cite{Sikivie:1983ip,Raffelt:1996wa, Giannotti:2017hny}. Searches were carried out by several experimental collaborations by looking for $a \to \gamma$ Primakoff conversion in solid crystal detectors, including DAMA~\cite{Bernabei:2004fi} (NaI), CUORE~\cite{Li:2015tyq, Li:2015tsa} (TeO$_2$), Edelweiss-II~\cite{Armengaud_2013}, SOLAX~\cite{PhysRevLett.81.5068}, COSME~\cite{COSME:2001jci}, CDMS~\cite{CDMS:2009fba}, and Majorana~\cite{Majorana:2022bse} (Ge). Other upcoming experiments like SuperCDMS~\cite{SuperCDMS:2022kse}, LEGEND~\cite{LEGEND:2021bnm}, and SABRE~\cite{SABRE:2018lfp} are projected to greatly expand coverage over the axion parameter space and test QCD axion solutions to the strong CP problem in the eV mass range. These experiments aim to take advantage of coherency in the conversion rate when axions satisfy the Bragg condition, enhancing the detection sensitivity by orders of magnitude relative to incoherent scattering.

Searching for solar axions via their coherent conversion in perfect crystals was first treated by Buchmuller \& Hoogeveen~\cite{Buchmuller:1989rb} using the Darwin theory of classical x-ray diffraction under the Bragg condition~\cite{warren}. The authors also alluded to potential enhancements in the signal yield when one considers the symmetrical Laue-case of diffraction for the incoming ALP waves. Yamaji \textit{et al}~\cite{Yamaji:2017pep} treated this case thoroughly for the 220 plane of cubic crystals, also using the classical theory, and included the effect of anomalous absorption, also known as the Borrmann effect. It was shown by these authors that an enhancement to the signal yield was possible, replacing the Bragg penetration depth ($L_\textrm{bragg} \sim 1$ $\mu$m) with the Borrmann-enhanced attenuation length (ranging from 10$\mu$m all the way to centimeter scales). 

The effect of anomalous absorption of x-rays was first shown by Borrmann~\cite{borrmann1954}, theoretically explained by Zachariasen~\cite{zachariasenBook,zachariasenPNAS} and others later (Battermann~\cite{batterman1,batterman2}, Hirsch~\cite{Hirsch:a00588}). A quantum mechanical treatment was offered by Biagini~\cite{PhysRevA.42.3695,PhysRevA.44.645} in which the Borrmann effect was explained by means of a statistical treatment of the $\ket{\alpha}$ and $\ket{\beta}$ Bloch waves and their mutual interference at the occupation number level. More recently, another treatment of the Borrmann effect in the quantum limit was applied to study photon-photon dissipation on Bragg-spaced arrays of superconduncting qubits~\cite{Poshakinskiy_2021}.

Now, the calculation of the event rates expected for the Primakoff conversion of solar axions coherently with a perfect crystal was treated in a more traditional, particle physics-based approach in refs.~\cite{Cebrian:1998mu,Bernabei:2001ny,Li:2015tsa} and it was applied to derive many of the constraints set by crystal-based solar axion experiments including DAMA, CUORE, Edelweiss-II, SOLAX, COSME, CDMS, and Majorana Demonstrator~\cite{Bernabei:2004fi,Li:2015tyq,Armengaud_2013,PhysRevLett.81.5068,COSME:2001jci,CDMS:2009fba,Majorana:2022bse}. However, absorption effects in Bragg and Laue case diffraction were not considered in refs.~\cite{Cebrian:1998mu,Bernabei:2001ny,Li:2015tsa}; indeed, when comparing the event rates between these references and those presented in light-shining-through-wall (LSW) experiments, which used the classical Darwin theory approach (e.g. ref.~\cite{Buchmuller:1989rb} and more recently ref.~\cite{Yamaji:2017pep}), there is a clear inconsistency. While the event rates in the LSW literature only consider the coherent volume of the crystal up to the relevant attenuation length ($\lambda \sim 1$ $\mu$m in the Bragg diffraction case or $\lambda \lesssim 100$ $\mu$m in the Laue-case), the solar axion searches have considered the whole volume of the crystal to exhibit coherency. In this work, we show that such effects reduce the expected event rates potentially up to the $\mathcal{O}(10^3)$ level depending on the assumed crystal size (and therefore, the assumed coherent volume enhancement) and material. Although this may impact the existing sensitivities set by solar axion searches in solid crystals, measures can be taken to optimize suppression of the event rate due to absorption effects and recover some or potentially all of the coherent volume.

In this section, I will re-derive the event rate formula for solar axion Primakoff scattering under the Bragg condition with full volume coherence. I will then motivate the inclusion of absorption effects and discuss the anomalous enhancement to the absorption length under the Borrmann effect. I will then write down the event rates for a perfect crystal exposed to the solar axion flux with and without the absorption effects and discuss the relevant phenomenology to set up the subsequent chapter

\section{Crystal structure}
For crystals with diamond structure (Ge, Si, C), we may have the diamond cubic, also seen as the inter-penetrating face-centered cubic (FCC), as the unit cell;
\begin{figure}[h]
    \centering
    \includegraphics[width=0.6\textwidth]{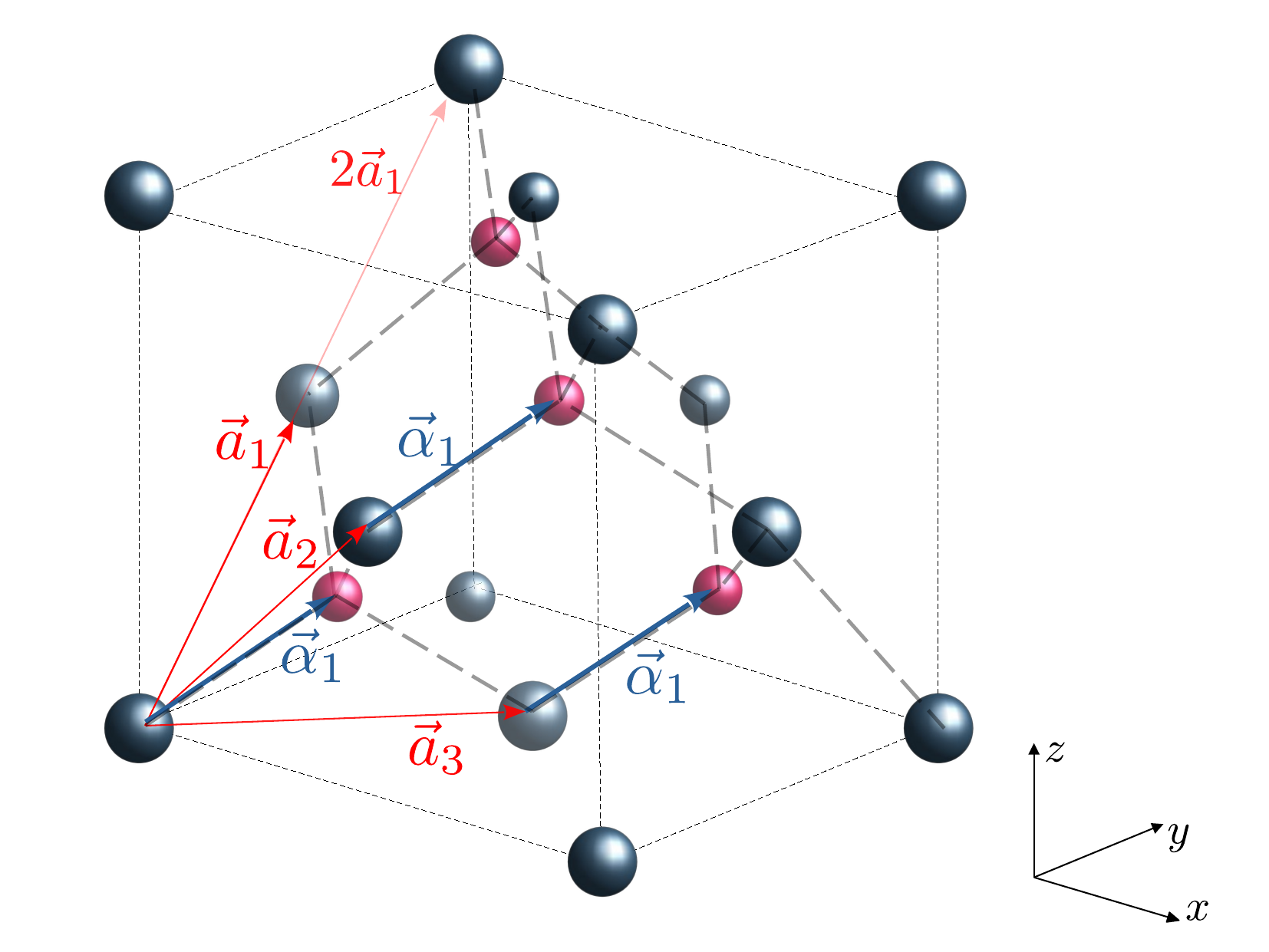}
    \caption{Unit cell for a diamond cubic or interfaced-FCC lattice, where the Bravais basis vectors $\vec{a}_i$ and primitive basis vectors $\vec{\alpha}_j$ are shown. In the case of FCC lattices like those for NaI or CsI crystals, the Na/Cs atoms are shown in blue and the I atoms shown in red, while for diamond cubics like C, Ge, or Si, each scattering center is an identical atomic species.}
    \label{fig:cubic}
\end{figure}
The \textit{primitive basis vectors} just describe the first two atoms on the bottom-left of Figure~\ref{fig:cubic}, left; these are
\begin{align}
    \vec{\alpha}_0 &= (0,0,0) \nonumber \\
    \vec{\alpha}_1 &= \frac{a}{4} (1,1,1)
\end{align}
while the \textit{basis vectors of the Bravais lattice} are described by $\vec{a}_1$, $\vec{a}_2$, and $\vec{a}_3$ in Figure~\ref{fig:cubic}, right:
\begin{align}
    \vec{a}_1 &= \frac{a}{2}(0,1,1) \nonumber \\
    \vec{a}_2 &= \frac{a}{2} (1,0,1) \nonumber \\
    \vec{a}_3 &= \frac{a}{2} (1,1,0)
\end{align}
we can translate anywhere on the lattice by stepping in integer multiples of these basis vectors;
\begin{equation}
    \vec{r} = n_1 \vec{a}_1 + n_2 \vec{a}_2 + n_3 \vec{a}_3
\end{equation}
We can then introduce the reciprocal lattice, which is like the Fourier transform of the real space lattice basis vectors. They reciprocal lattice basis vectors $\vec{b}_i$ satisfy $\vec{b}_i \cdot \vec{a}_j = 2\pi \delta_{ij}$. In general the transformations give
\begin{align}
\vec{b}_1 &= 2\pi \dfrac{\vec{a}_2 \times \vec{a}_3}{|\vec{a}_1 \cdot (\vec{a}_2 \times \vec{a}_3)|} \nonumber \\
\vec{b}_2 &= 2\pi \dfrac{\vec{a}_3 \times \vec{a}_1}{|\vec{a}_1 \cdot (\vec{a}_2 \times \vec{a}_3)|} \nonumber \\
\vec{b}_3 &= 2\pi \dfrac{\vec{a}_1 \times \vec{a}_2}{|\vec{a}_1 \cdot (\vec{a}_2 \times \vec{a}_3)|}
\end{align}
for our diamond cubic, we have
\begin{align}
    \vec{b}_1 &= \dfrac{2\pi}{a} (-1, 1, 1) \nonumber \\
\vec{b}_2 &= \dfrac{2\pi}{a} (1, -1, 1) \nonumber \\
\vec{b}_3 &= \dfrac{2\pi}{a} (1, 1, -1)
\end{align}
One can check again in the figure that in real space, the reciprocal basis vectors correspond to momentum (Bragg condition) that point along the surface normals of the surfaces defined by the lattice basis vectors; the reciprocal vectors are momenta that point normal to the lattice planes. Now use these basis vectors to write any reciprocal lattice vector in terms of Miller indices $h,k,l \in \mathbb{Z}$;
\begin{equation}
    \vec{G} = h \vec{b}_1 + k \vec{b}_2 + l \vec{b}_3
\end{equation}
Sometimes the integers $h,k,l$ are used instead, and in some contexts one can use this basis to express $\vec{G}$ as
\begin{equation}
    \vec{G}(hkl) = \dfrac{2\pi}{a} (h, k, l)
\end{equation}
The lattice constants, cell volumes, and basis vectors for a few examples (Ge, Si, CsI, and NaI) are listed in Table~\ref{tab:lattice}.

\begin{table*}[]
\begingroup
    \centering
    \renewcommand*{\arraystretch}{1.2}
    \begin{tabular}{| c | c | c | c | c | c |}
    \hline
        Material & \thead{Lattice Constant\\$a$ (\AA)} & \thead{Cell Volume\\ $v_\textrm{cell}$ (\AA$^3$)} & \thead{Primitive Basis} & \thead{Bravais Basis} & $\vec{G}\times a / 2\pi$\\
        \hline
         Ge (Diamond Cubic) & 5.657 & 181.0 & \multirow{2}{*}{\thead{$\vec{\alpha}_0 = (0,0,0)$\\$\vec{\alpha}_1 = \frac{a}{4}(1,1,1)$}} & \multirow{4}{*}{\thead{$\vec{a}_1 = \frac{a}{2}(1,0,1)$\\$\vec{a}_2 = \frac{a}{2}(0,1,1)$\\$\vec{a}_3 = \frac{a}{2}(1,1,0)$}} & \multirow{4}{*}{\thead{$(m_1 - m_2 + m_3,$\\$-m_1 + m_2 + m_3,$\\$m_1 + m_2 - m_3)$}}\\
         Si (Diamond Cubic) & 5.429 & 160.0 & & & \\
         \cline{1-4}
         CsI (FCC) & 4.503 & 91.3 & \multirow{2}{*}{\thead{$\vec{\alpha}_0 = (0,0,0)$\\$\vec{\alpha}_1 = \frac{a}{2}(1,1,1)$}} & & \\
         NaI (FCC) & 6.462 & 67.71 & & & \\
         \hline
    \end{tabular}
\endgroup
    \caption{Lattice information for typical crystal detector technologies, FCC and diamond cubic.}
    \label{tab:lattice}
\end{table*}
\begin{table}[]
    \centering
    \begin{tabular}{|c|c|c|c|c|}
\hline
$hkl$	& $\sin\theta/ \lambda$	& $|S(hkl)|$ & $1/|G|^2$ & $(|S(hkl)|/|\vec{G}|)^2$ \\
\hline
111	&	0.1530891645	&	5.7	&	0.2702037463	&	8.778919716	\\
022	&	0.3535442814	&	8	&	0.05066320242	&	3.242444955	\\
202	&	0.3535442814	&	8	&	0.05066320242	&	3.242444955	\\
220	&	0.3535442814	&	8	&	0.05066320242	&	3.242444955	\\
113	&	0.3852659487	&	5.7	&	0.04266374941	&	1.386145218	\\
131	&	0.3852659487	&	5.7	&	0.04266374941	&	1.386145218	\\
311	&	0.3852659487	&	5.7	&	0.04266374941	&	1.386145218	\\
133	&	0.4592674936	&	5.7	&	0.03002263847	&	0.975435524	\\
313	&	0.4592674936	&	5.7	&	0.03002263847	&	0.975435524	\\
331	&	0.4592674936	&	5.7	&	0.03002263847	&	0.975435524	\\
333	&	0.4592674936	&	5.7	&	0.03002263847	&	0.975435524	\\
224	&	0.4999871177	&	8	&	0.02533160121	&	1.621222478	\\
242	&	0.4999871177	&	8	&	0.02533160121	&	1.621222478	\\
422	&	0.4999871177	&	8	&	0.02533160121	&	1.621222478	\\
004	&	0.6123566581	&	8	&	0.01688773414	&	1.080814985	\\
040	&	0.6123566581	&	8	&	0.01688773414	&	1.080814985	\\
400	&	0.6123566581	&	8	&	0.01688773414	&	1.080814985	\\
444	&	0.6123566581	&	8	&	0.01688773414	&	1.080814985	\\
335	&	0.6312027955	&	5.7	&	0.01589433802	&	0.5164070421	\\
353	&	0.6312027955	&	5.7	&	0.01589433802	&	0.5164070421	\\
533	&	0.6312027955	&	5.7	&	0.01589433802	&	0.5164070421	\\
115	&	0.6789062885	&	5.7	&	0.01373917354	&	0.4463857483	\\
135	&	0.6789062885	&	5.7	&	0.01373917354	&	0.4463857483	\\
151	&	0.6789062885	&	5.7	&	0.01373917354	&	0.4463857483	\\
153	&	0.6789062885	&	5.7	&	0.01373917354	&	0.4463857483	\\
315	&	0.6789062885	&	5.7	&	0.01373917354	&	0.4463857483	\\
351	&	0.6789062885	&	5.7	&	0.01373917354	&	0.4463857483	\\
511	&	0.6789062885	&	5.7	&	0.01373917354	&	0.4463857483	\\
513	&	0.6789062885	&	5.7	&	0.01373917354	&	0.4463857483	\\
531	&	0.6789062885	&	5.7	&	0.01373917354	&	0.4463857483	\\
044	&	0.7070885628	&	8	&	0.01266580061	&	0.8106112388	\\
404	&	0.7070885628	&	8	&	0.01266580061	&	0.8106112388	\\
440	&	0.7070885628	&	8	&	0.01266580061	&	0.8106112388	\\
355	&	0.723471166	&	5.7	&	0.01209867521	&	0.3930859574	\\
535	&	0.723471166	&	5.7	&	0.01209867521	&	0.3930859574	\\
553	&	0.723471166	&	5.7	&	0.01209867521	&	0.3930859574	\\
555	&	0.7654458227	&	5.7	&	0.01080814985	&	0.3511567886	\\
\hline
    \end{tabular}
    \caption{Example values of $\sin\theta/\lambda$ for a diamond cubic lattice (Ge) in terms of Miller indices $hkl$ and the structure factor that appears in the event rate.}
    \label{tab:miller_table}
\end{table}

\section{Laue and Bragg Conditions}
For a real space coordinate $\vec{x}$, the Laue condition for diffraction reads
\begin{equation}
    \exp (i \vec{x}\cdot \vec{G}) = 1
\end{equation}
or that $\vec{x}\cdot\vec{G} = 2\pi N$ for $N\in \mathbb{Z}$. This implies that $\vec{q} = \vec{G}$.
\begin{align}
    \text{The Laue Condition:} & \, \, \,  \vec{q} = \vec{G}
    \label{eq:laue}
\end{align}
A special case of the Laue condition is the Bragg condition, and it comes from imposing the elastic scattering condition. Rewrite Eq.~\ref{eq:laue} as follows;
\begin{align}
\label{eq:laue_momenta}
    \vec{G} &= \vec{q} = \vec{k}_f - \vec{k}_i \nonumber \\
    \vec{k}_i &= \vec{k}_f - \vec{G} \nonumber \\
    |\vec{k}_i|^2 &= |\vec{k}_f - \vec{G}|^2 \nonumber \\
    |\vec{k}_i|^2 &= |\vec{k}_f|^2 + |\vec{G}|^2 - 2 \vec{k}_f \cdot \vec{G}
\end{align}
The \textbf{Bragg condition} is a statement that we are in the elastic scattering regime; $|\vec{k}_i| = |\vec{k}_f|$. From Eq.~\ref{eq:laue_momenta} this implies
\begin{align}
    \text{The Bragg Condition (Version I)} & \, \, \, 2 \vec{k}_f \cdot \vec{G} = |\vec{G}|^2
    \label{eq:bragg1}
\end{align}
This can be recast in its more familiar form, taking $\sin\theta$ as the sine of the angle between $\vec{G}$ and $\vec{k}_f$, $|\vec{k}_f| = 2\pi/\lambda$, and $|\vec{G}| = 2\pi n/d$:
\begin{align}
    \text{The Bragg Condition (Version II)} & \, \, \, 2d \sin\theta = n\lambda
\end{align}
Lastly, there is one more way to represent the Bragg condition which has shown up in various ALP-crystal scattering papers, and it follows easily from Eq.~\ref{eq:bragg1}:
\begin{align}
    \text{The Bragg Condition (Version III)} & \, \, \, E_i = \dfrac{|\vec{G}|^2}{2 \hat{u}\cdot\vec{G}}
    \label{eq:bragg3}
\end{align}
It should be noted that \textbf{the Bragg and Laue conditions are different than Bragg-case and Laue-case diffraction}. The terms ``Bragg-case" and ``Laue-case" diffraction (or conversion in the case of ALP-photon interactions) have to do with the orientation of the crystal geometry with respect to the incident wave.

\begin{figure}[h]
    \centering
    \includegraphics[width=0.32\textwidth]{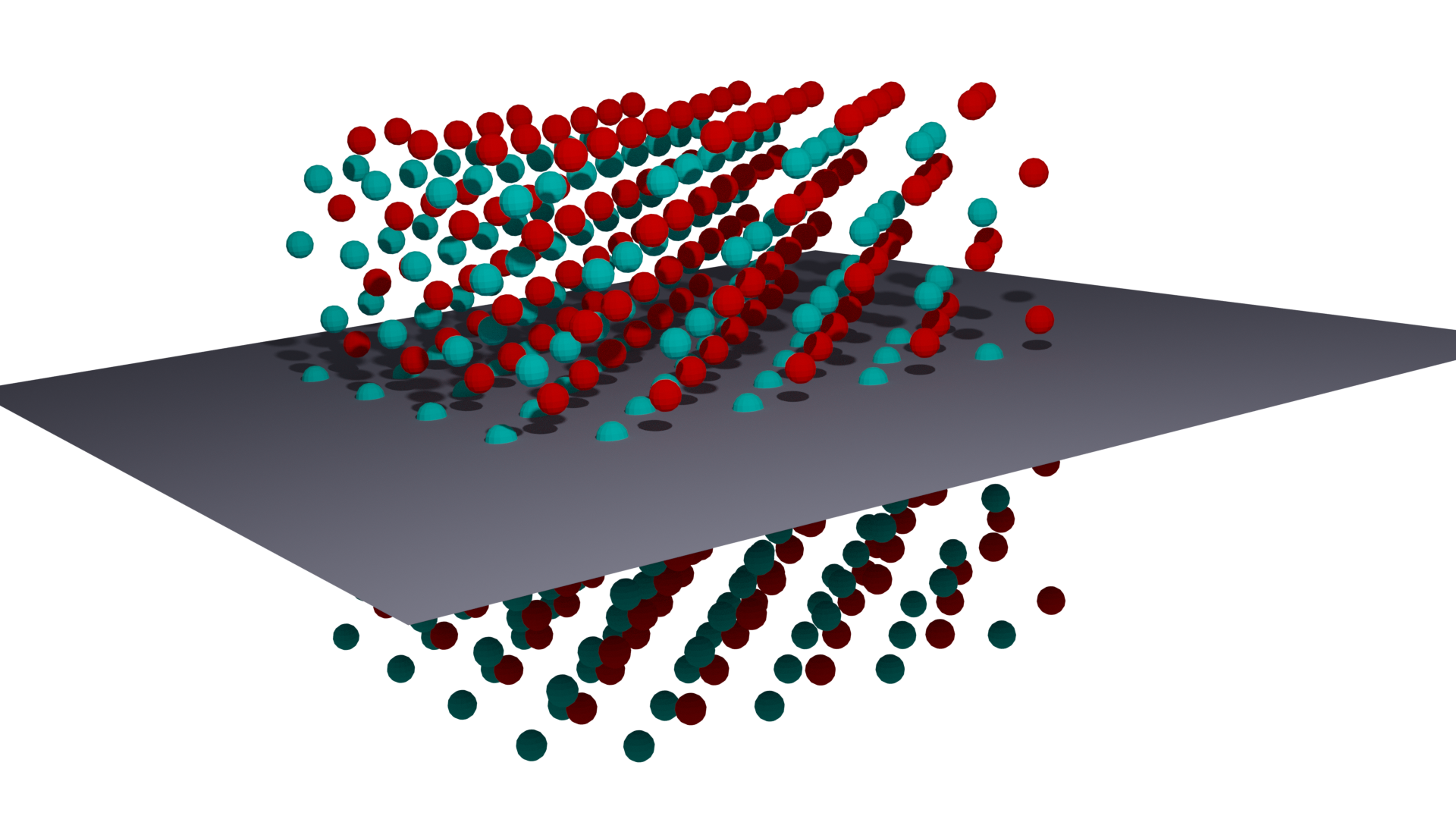}
    \includegraphics[width=0.32\textwidth]{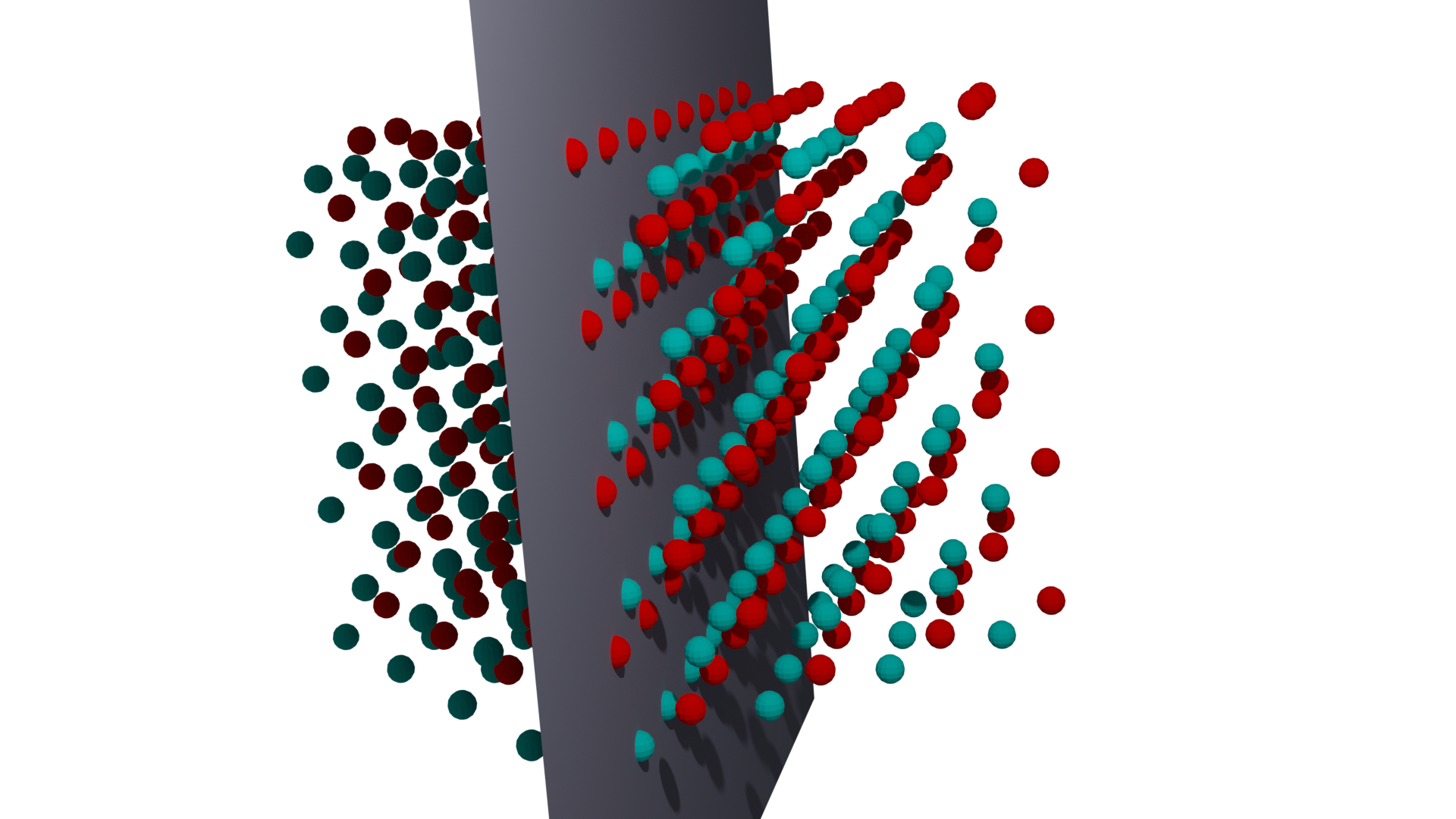}
    \includegraphics[width=0.32\textwidth]{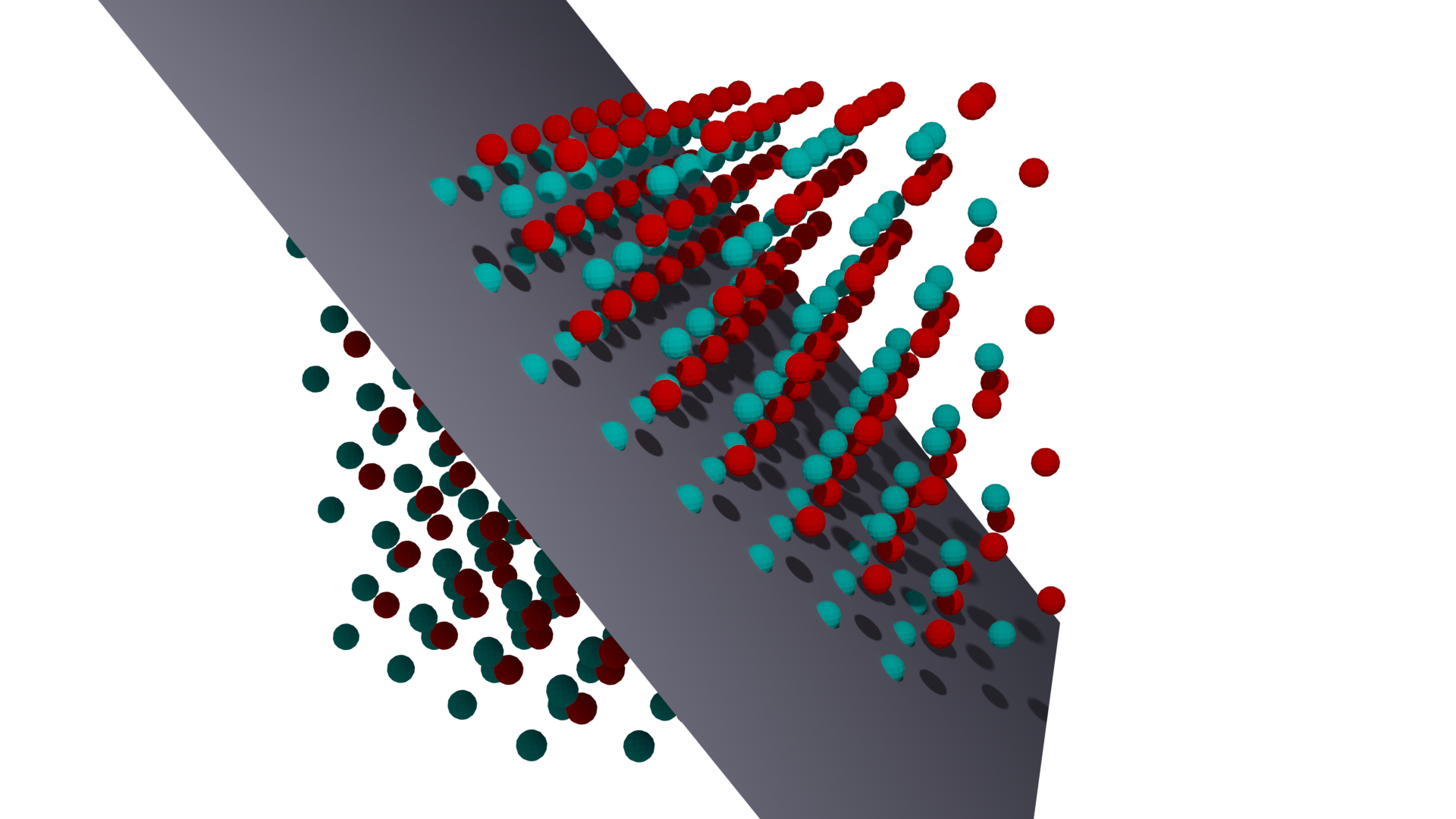}
    \caption{Crystallographic planes cut by the $hkl=220$ (left), 224 (middle), and 333 (right) reciprocal lattice vectors in CsI.}
    \label{fig:csi_planes}
\end{figure}

\section{Event Rates for Coherent Bragg-Primakoff Scattering from Solar Axions}

Let $f(\vec{k},\vec{k}^\prime)$ be the Primakoff scattering matrix element for a single atomic target, for an incoming ALP 3-momentum $\vec{k}$ and outgoing $\gamma$ 3-momentum $\vec{k}^\prime$;
\begin{equation}
    f = \braket{\mathcal{M}_\textrm{free}} F_A (q)
\end{equation}
where $F_A$ is the atomic form factor given in Eq.~\ref{eq:atomic_ff} and $\mathcal{M}_\textrm{free}$ is the single-atomic scattering amplitude, Eq.~\ref{eq:primakoff_m2} taken in the relativistic limit
\begin{equation}
\label{eq:prim_rel}
    \braket{\mathcal{M}_\textrm{free}} = \dfrac{4 e^2 g_{a\gamma}^2}{t^2} E_\gamma^2 m_N^2 k^2 \sin^2\theta 
\end{equation}
Similar to the approach illustrated by Bednyakov and Naumov to get the total coherent amplitude~\cite{Bednyakov:2021lul} for neutrinos scattering over $N$ scattering centers in a nucleus, we sum over the $N$ scattering centers in a crystal;
\begin{equation}
    \mathcal{M}(\vec{k},\vec{k}^\prime) = \sum_{j=1}^N f_j(\vec{k},\vec{k}^\prime) e^{i(\vec{k}^\prime - \vec{k})\cdot \vec{r}_j}
\end{equation}
where $e^{i(\vec{k}^\prime - \vec{k})\cdot \vec{r}_j}$ is a phase factor that comes from assuming plane wave solutions for the in and out states. The position vector $\vec{r}_j$ can be expressed in terms of the Bravais lattice basis vectors and the primitive basis vectors for each unit cell of the crystal. For germanium crystal with lattice constant $a$, we have \textit{primitive basis vectors} $\alpha_j$ while the Bravais lattice vectors are described by $\vec{a}_1$, $\vec{a}_2$, and $\vec{a}_3$. We can then represent any scattering site as a linear combination of the $a$'s and either the first or second primitive;
\begin{align}
    \vec{r}_{i,0} &= \vec{R}_i  + \vec{\alpha}_0 = n_1 \vec{a}_1 + n_2 \vec{a}_2 + n_3 \vec{a}_3 + \vec{\alpha}_0 \\
    \vec{r}_{i,1} &= \vec{R}_i  + \vec{\alpha}_1 = n_1 \vec{a}_1 + n_2 \vec{a}_2 + n_3 \vec{a}_3 + \vec{\alpha}_1
\end{align}
where the index $i$ maps to a unique combination ($n_1, n_2, n_3$).
If we square this, we get
\begin{equation}
     \mid\mathcal{M}(\vec{k},\vec{k}^\prime)\mid^2 = \sum_{i=1}^N \mid f_i\mid^2 + \sum_{j\neq i}^N \sum_{i=1}^N f_j^\dagger f_i e^{-i\vec{q}\cdot(\vec{r}_i - \vec{r}_j)}
\end{equation}
taking $\vec{q} \equiv \vec{k} - \vec{k}^\prime$. Rewriting in terms of a sum over $N_c$ cells and the cell primitives, the coherent part (second term) is
\begin{equation}
     \mid\mathcal{M}(\vec{k},\vec{k}^\prime)\mid^2 = \sum_{j\neq i}^{N_c} \sum_{i=1}^{N_c} \sum_{\mu = 0}^1\sum_{\nu = 0}^1 f_j^\dagger f_i e^{-i\vec{q}\cdot(\vec{R}_i - \vec{R}_j + \vec{\alpha}_\mu - \vec{\alpha}_\nu)}
\end{equation}
When the Laue condition is met, we have $\vec{q} = \vec{G}$ and $\vec{G}\cdot \vec{R}_i$ is a $2\pi$ integer multiple;
\begin{equation}
 |\mathcal{M}|^2 \equiv \sum_{j\neq i}^{N_c} \sum_{i=1}^{N_c} \sum_{\mu,\nu = 0}^1 f_j^\dagger f_i e^{-i\vec{G}\cdot(\vec{\alpha}_\mu - \vec{\alpha}_\nu)}
\end{equation}
Now we can factorize the sum over primitives, and since we are considering a monoatomic crystal we can also take the $f_i = f_j$, simplifying things;
\begin{equation}
 |\mathcal{M}|^2 = N_c^2 f^\dagger f \sum_{\mu,\nu = 0}^1  e^{-i\vec{G}\cdot(\vec{\alpha}_\mu - \vec{\alpha}_\nu)} 
    \label{eq:cohsum}
\end{equation}
Now the structure function is nothing but the sum over primtives;
\begin{equation}
    S(\vec{G}) = \sum_\mu e^{i \vec{G}\cdot\alpha_\mu}
\end{equation}
and we have no need for a species index $j$ on $S_j(\vec{G})$ since we only have one atomic species, but it is trivial to extend this derivation to include it - we just need to add another index to the primitive basis vectors and sum over it. With this identification and also taking $f^\dagger f = |\mathcal{M}_\textrm{free}|^2 F^2_A (\vec{G})$, we have
\begin{align}
 |\mathcal{M}|^2 &= N_c^2 |\mathcal{M}_\textrm{free}|^2 |F_A (\vec{G}) S(\vec{G})|^2
\end{align}

Now let's write down the cross section. From the Lorentz-invariant phase space element we have
\begin{equation}
    d\sigma = \dfrac{1}{4 E_a m_N v_a} |\mathcal{M}|^2 \dfrac{d^3 k^\prime}{(2\pi)^3 2E_\gamma}\dfrac{d^3 p^\prime}{(2\pi)^3 2E_{p^\prime}} (2\pi)^4 \delta^4 (k + p - k^\prime - p^\prime)
\end{equation}
Taking the ALP velocity $v_a = 1$, momentum transfer minimal such that $E_{p^\prime} = m_N$, and integrating out the $\delta^3$ we get
\begin{equation}
    d\sigma = \dfrac{1}{64 \pi^2 E_a E_\gamma m_N^2} |\mathcal{M}|^2  d^3 k^\prime \delta(E_a - E_\gamma)
\end{equation}
Performing a change of variables to $d^3k^\prime \to d^3q$ (since $q = k - k^\prime$ and $k$ is fixed), we would integrate this over $\vec{q}$. Since we have $\vec{q} = \vec{G}$ at this stage, we should replace the integral with a sum;
\begin{equation}
    \int d^3 q \to \frac{(2\pi)^3}{V} \sum_{\vec{G}}
\end{equation}
The event rate formula is constructed from a convolution of the detector response (taken as a gaussian smearing function between true energy $E_\gamma$ and electron-equivalent energy $E_{ee}$, with width $\Delta \sim \mathcal{O}(1)$ keV), axion flux $\Phi_a$, and cross section;
\begin{equation}
    R = \int_{E_1}^{E_2} dE_{ee} \int_0^\infty dE_a  \frac{(2\pi)^3}{V}\sum_{\vec{G}} \dfrac{d\Phi_a}{d E_a}   \dfrac{1}{64 \pi^2 E_a E_\gamma m_N^2} |\mathcal{M}|^2 \delta(E_a - E_\gamma) \cdot \bigg( \dfrac{1}{\Delta \sqrt{2\pi}} e^{-(E_{ee} - E_\gamma)^2/2\Delta^2} \bigg)
\end{equation}
Putting in the definition of $|\mathcal{M}|^2$ that we worked out and substituting the free Primakoff cross section in the relativistic limit (Eq.~\ref{eq:prim_rel}), integrating over the energy delta function (and identifying $E_a = E_\gamma = k$ for simplicity), and integrating over $dE_{ee}$ we arrive finally at the solar axion event rate formula under full volume coherence,
\begin{equation}
\label{eq:event_rate_app}
\boxed{
    R = \dfrac{(2\pi)^3 e^2 g_{a\gamma}^2}{8 \pi^2} \dfrac{V}{v_c^2} \sum_{\vec{G}} \dfrac{d\Phi_a}{dE} \dfrac{k^2 \sin^2 (2\theta)}{|\vec{G}|^4} |F_A(\vec{G})S(\vec{G})|^2 \mathcal{W}(E_1, E_2, E)}
\end{equation}
The event rate in Eq.~\ref{eq:event_rate_app} encodes the effect of detector energy resolution $\Delta$ within the function $\mathcal{W}$ resulting from the $dE_{ee}$ integral;
\begin{equation}
    \mathcal{W}(E_a, E_1, E_2, \Delta) = \frac{1}{2}\bigg(\textrm{erf}\bigg(\frac{E_a - E_1}{\sqrt{2}\Delta}\bigg) - \textrm{erf}\bigg(\frac{E_a - E_2}{\sqrt{2}\Delta}\bigg) \bigg)
\end{equation}

One can see the tabulated values of the structure factor and corresponding miller indices $h,k,l$ in Table~\ref{tab:miller_table}, as well as the decreasing values of $(S(h,k,l)/|\vec{G}|)^2$ as they appear in the $\vec{G}$ sum, shown there for a Ge diamond cubic lattice for reference.

Eq.~\ref{eq:event_rate_app} is almost identical to the rate in ref.~\cite{Cebrian:1998mu}, which uses a different definition of the atomic form factor up to a factor of $\frac{q^2}{e k^2}$. After some algebra, the event rate in Eq.~\ref{eq:event_rate_app} is still different than that given in ref.~\cite{Cebrian:1998mu} up to a factor of $4\sin^2(\theta) \sim \mathcal{O}(1)$, although the event rate formula derived here is consistent with the calculation performed in refs.~\cite{Li:2015tsa, Li:2015tyq}.

However, the most important aspect that has been neglected up until now is the attenuation of the final state photon in dielectric that will alter and suppress the coherence that we have relied on so far. In this next section I will address this issue.

\section{Coherence and Absorption}
Let $f(\vec{k},\vec{k}^\prime)$ be the Primakoff scattering matrix element for a single atomic target, for an incoming ALP 3-momentum $\vec{k}$ and outgoing $\gamma$ 3-momentum $\vec{k}^\prime$. Similar to the approach illustrated by Bednyakov and Naumov to get the total coherent amplitude~\cite{Bednyakov:2021lul} for neutrinos scattering over $N$ scattering centers in a nucleus, we sum over the $N$ scattering centers in a crystal;
\begin{equation}
    \mathcal{M}(\vec{k},\vec{k}^\prime) = \sum_{j=1}^N f_j(\vec{k},\vec{k}^\prime) e^{i(\vec{k}^\prime - \vec{k})\cdot \vec{r}_j}
\end{equation}
where $e^{i(\vec{k}^\prime - \vec{k})\cdot \vec{r}_j}$ is a phase factor that comes from assuming plane wave solutions for the in and out states. If we square this, we get
\begin{equation}
     \mid\mathcal{M}(\vec{k},\vec{k}^\prime)\mid^2 = \sum_{i=1}^N \mid f_i\mid^2 + \sum_{j\neq i}^N \sum_{i=1}^N f_j^\dagger f_j e^{-i\vec{q}\cdot(\vec{r}_i - \vec{r}_j)}
\end{equation}
taking $\vec{q} \equiv \vec{k} - \vec{k}^\prime$. Demanding the Laue diffraction condition, $\vec{q}\cdot(\vec{r}_i - \vec{r}_j) = 2\pi n$ for $n\in \mathbb{Z}$, then the phase factor in the exponential goes to one. In this limit, the diagonal (first) term is subdominant and the final matrix element squared tends to $\mathcal{M}^2 \to N^2 f^2$.

Now consider interactions of the final state $\gamma$ with the crystal lattice, including the absorption and scattering effects. Pragmatically, we modify the plane wave solutions of the final state photon to that of one in a dielectric medium,
\begin{equation}
    \vec{k}^\prime \to \Bar{n} \vec{k}^\prime, \, \, \, \Bar{n} = n - i \kappa,
    \label{eq:dielectric}
\end{equation}
where $\bar{n}$ is the complex index of refraction. Making this modification, we have
\begin{align}
    e^{i \bar{n} \vec{k}^\prime \cdot (\vec{r}_i - \vec{r}_j)} &\to e^{i n \vec{k}^\prime \cdot (\vec{r}_i - \vec{r}_j)} e^{-\frac{\mu}{2} |\hat{k}^\prime \cdot (\vec{r}_j - \vec{r}_i)|}
\end{align}
In the last line above, the absorption factor $\mu \equiv 2 \kappa |\vec{k}|$ makes its appearance. Equivalently, we can write the standard absorption coefficient as the product of the total photon absorption cross section multiplied by number density of the material in transport, $\mu= n \sigma$, and relates to the mean free path of the photon $\lambda \equiv 1 / \mu$. Conceptually, this factor encodes the effect of a reduced coherent interference amplitude between any two scattering centers, since a photon plane wave originated at one scattering center will have been attenuated after reaching another scattering center.

We note that Eq.~\ref{eq:dielectric} is a heuristic choice, since the attenuated plane wave solution is not a true eigenstate of the interaction Hamiltonian, but rather a simple ansatz made to capture the phenomenology of absorption. For further convenience we use $z_{ij} \equiv \mid\hat{k^\prime}\cdot(\vec{r}_i-\vec{r}_j)\mid$ and $\lambda = 1/\mu$, taking the real part of the index of refraction $\simeq 1$. We then have
\begin{equation}
     \mid\mathcal{M}(\vec{k},\vec{k}^\prime)\mid^2 = \sum_{i=1}^N \mid f_i\mid^2 + \sum_{j\neq i}^N \sum_{i=1}^N f_j^\dagger f_j e^{-i\vec{q}\cdot(\vec{r}_i - \vec{r}_j)}e^{-z_{ij}/(2\lambda)}
\end{equation}
At this stage evaluating the sum is tricky, so I will first give a rough estimation for the suppression factor of the modified coherent sum. Again, if we apply the Laue diffraction condition $\vec{q}\cdot(\vec{r}_i - \vec{r}_j) = 2\pi n$ and if the scattering centers are all identical ($f_j = f_i$), then 
\begin{align}
    \mid\mathcal{M}(\vec{k},\vec{k}^\prime)\mid^2 &\simeq f^\dagger f \sum_{j\neq i}^N \sum_{i=1}^N e^{-z_{ij}/(2\lambda)} \nonumber \\
    &=f^\dagger f \sum_{j\neq i}^N \int_\textrm{Crystal} d^3r \, e^{-z_{j}/(2\lambda)} \sum_{i=1}^N \delta^3(\vec{r} - \vec{r}_i)
\end{align}
Then going to the continuum limit, we have
\begin{align}
    \mid\mathcal{M}(\vec{k},\vec{k}^\prime)\mid^2 &\simeq f^\dagger f \sum_{j\neq i}^N  \dfrac{N}{V} \int d^3r \, e^{-z_{j}/(2\lambda)} \nonumber \\
    &\simeq f^\dagger f \sum_{j\neq i}^N  \dfrac{N}{V} \int_0^{L_x} \int_0^{L_y} \int_0^{L_z} dx dy dz \, e^{-|z-\hat{z}\cdot\vec{r}_j|/(2\lambda)} \nonumber \\
    &\simeq f^\dagger f \dfrac{ L_x L_y N}{V} \sum_{j\neq i}^N   \int_0^{L_z}dz \, e^{-|z-\hat{z}\cdot\vec{r}_j|/(2\lambda)}
    \label{eq:abs_volume}
\end{align}
Above, we used the fact that for Primakoff forward scattering, $\hat{k} \simeq \hat{k}^\prime$, so $z_{ij} \simeq \mid\hat{z}\cdot(\vec{r}_i-\vec{r}_j)\mid$, choosing the direction of the incoming ALP momentum $\hat{k} = \hat{z}$ without loss of generality.

Let us use the definition $z_0 \equiv \hat{z}\cdot\vec{r}_j$, then we see
\begin{eqnarray}
&&\int_{0}^{L_z}e^{-|z-\hat{z}\cdot\vec{r}_j|/(2\lambda)}dz = \int_{0}^{L_z}e^{-|z-z_0|/(2\lambda)} 
\\
&=& \int_{0}^{z_0} e^{(z-z_0)/(2\lambda)}dz + \int_{z_0}^{L_z} e^{-(z-z_0)/(2\lambda)}dz
\\
&=&\lambda - \frac{\lambda}{2}\left(e^{-z_0/(2\lambda)} + e^{-(L_z-z_0)/(2\lambda)}\right)
\label{eq:integral}
\end{eqnarray}
For $L_z\gg\lambda$, this result varies from $\lambda$ at $z_0 = 0$ to a max of $2\lambda$ at $z_0 = L_z/2$, and back to $\lambda$ at $z_0 = L_z$. Taking the sum $\sum_{j\neq i}^N$ to the continuum limit, one finds that the term in parentheses in Eq.~\ref{eq:integral} is $\mathcal{O}(\lambda^2) << L_z$.
Therefore, we have
\begin{align}
    \mid\mathcal{M}(\vec{k},\vec{k}^\prime)\mid^2 &\gtrsim f^\dagger f\sum_{j\neq i}^N \dfrac{\lambda L_x L_y N}{V} \nonumber \\
    &\gtrsim f^\dagger f N^2 \dfrac{\lambda}{L_z}
\end{align}
Comparing Eq.~\ref{eq:abs_volume} to the usual result $\propto N^2$, we see that the coherent volume is $V \times \lambda / L_z$, and the total scattering rate is suppressed by a factor $\lambda/L_z$.

This inequality above is strictly a lower limit because, as we will show in \S~\ref{sec:borrmann}, the suppression to the coherent sum by the absorptive sum, which we label as $I$,
\begin{equation}
\label{eq:abs_sum}
   I \equiv \sum_{j\neq i}^N \sum_{i=1}^N e^{-z_{ij}/(2\lambda)},
\end{equation}
may be mitigated under certain conditions. Therefore, the suppression factor $\lambda / L_z$ serves as a pessimistic guiding estimate, but in principle we should compute the sum in Eq.~\ref{eq:abs_sum} explicitly. After rederiving the coherent sum using the replacements in Eqns.~\ref{eq:dielectric}, the event rate becomes
\begin{equation}
\label{eq:event_rate_app_abs}
    \boxed{\dfrac{dN}{dt} = \dfrac{(2\pi)^3 e^2 g_{a\gamma}^2}{8 \pi^2} \dfrac{V}{v_\text{cell}^2} \sum_{\vec{G}} I(\vec{k},\vec{G}) \dfrac{d\Phi_a}{dE} \dfrac{k^2 \sin^2 (2\theta)}{|\vec{G}|^4} |F_A(\vec{G})S(\vec{G})|^2 \mathcal{W}(E_1, E_2, E)}
\end{equation}

\section{The Borrmann Effect}
\label{sec:borrmann}
Yamaji \textit{et. al.}~\cite{Yamaji:2017pep} has found that for the Laue-case conversion of ALPs, the attenuation length is modified as
\begin{equation}
    L_\text{att} \to L_{\alpha / \beta} \equiv 2L_{\text{att},\alpha / \beta}\bigg(1 - \exp\bigg(-\dfrac{L}{2L_{\text{att},\alpha / \beta}} \bigg) \bigg) 
\end{equation}
where $L_{\text{att},\alpha / \beta} = \dfrac{L_\text{att}}{1 \mp \epsilon}$ and $\epsilon$ is a ratio involving the imaginary parts of the scattering form factor:
\begin{equation}
    \epsilon \equiv \dfrac{\text{Im}\{F(\vec{G})\}}{\text{Im}\{F(\vec{0})\}}
\end{equation}
These modifications come from the anomalous dispersion or anomalous absorption effect, or the Borrmann effect. It is an effect that occurs for so-called ``Bloch waves" $\alpha$ and $\beta$ that form in the crystal. The notes below will attempt to explain this further. \\

The total scattering form factor can be decomposed into the real and imaginary parts~\cite{chantler_NIST};
\begin{equation}
    f = f^0 + \Delta f^\prime + i \Delta f^{\prime\prime}
\end{equation}
where $f^0$ is the atomic form factor, usually given as the Fourier transform of the charge density;
\begin{equation}
    f^0(q) \equiv \int d^3 \vec{x} \rho(\vec{x}) e^{i \vec{q}\cdot\vec{x}}
\end{equation}
The second term in the real part of the form factor is the anomalous form factor $\Delta f^\prime$, and $\Delta f^{\prime\prime}$ is the imaginary part of the form factor associated with absorption. There is a useful relationship between the imaginary part and the photoelectric absorption cross section;
\begin{equation}
    \Delta f^{\prime\prime}(E) = \dfrac{E \sigma_\text{PE}(E)}{2\hbar c r_e}
    \label{eq:imff}
\end{equation}
The anomalous absorption due to the Borrmann effect modifies the absorption coefficient $\mu_0 = 1/\lambda$ as
\begin{equation}
    \frac{1}{\lambda_\text{eff}} = \mu_\text{eff} = \mu_0 \bigg[1 - \dfrac{F^{\prime\prime}(hkl)}{F^{\prime\prime}(000)}\bigg]
    \label{eq:anomalous_depth}
\end{equation}
Here $F^{\prime\prime}(hkl)$ is the combination of structure function and imaginary form factor, $F^{\prime\prime}(hkl) = S(hkl) \Delta f^{\prime\prime}$~\cite{batterman1, batterman2}. The ratio in the second term of the expression is usually denoted as $\epsilon$\footnote{In Yamaji et al~\cite{Yamaji:2017pep}, they use $\kappa$.}
\begin{align}
    \epsilon &\equiv \dfrac{F^{\prime\prime}(hkl)}{F^{\prime\prime}(000)} \nonumber \\
    &= \dfrac{e^{-M} \sum_j Z_j \eta_j f_j^0 (hkl)}{\sum_j Z_j \eta_j} \cdot \dfrac{S(h,k,l)}{S(0,0,0)} \nonumber \\
    &= \dfrac{e^{-M} \Delta f^{\prime\prime}(hkl)}{\sum_j Z_j \eta_j} \cdot \dfrac{S(h,k,l)}{S(0,0,0)}
\end{align}
The exponential $e^{-M}$ is a Debye-Waller factor to account for thermal effects and is usually close to 1.

Alternatively, we can calculate the imaginary form factor using the relation in Wagenfield's paper. Wagenfield's form factor for the anomalous dispersion of X-rays with incoming and outgoing momenta and polarizations $\vb{k}$, $\polvec_0$, $\vb{k}^\prime$, $\polvec_0^\prime$  is~\cite{wagenfield1986}
\begin{equation}
    \Delta f^{\prime\prime} = \dfrac{\pi \hbar^2}{m_e} \bigg( \int \psi_f^*(\vb{r}) \polvec_0 \cdot \vb{\nabla} e^{i \vb{k}\cdot\vb{r}} \psi_i(\vb{r}) d^3 r \bigg) \bigg( \int \psi_f(\vb{r}) \polvec^\prime_0 \cdot \vb{\nabla} e^{-i \vb{k}^\prime\cdot\vb{r}} \psi_i^*(\vb{r}) d^3 r \bigg)
\end{equation}
Applying the gradient and expanding, we get some terms proportional to $\polvec_0 \cdot \vb{k}$ which vanish, leaving us with
\begin{equation}
    \Delta f^{\prime\prime} = \dfrac{\pi \hbar^2}{m_e} \bigg( \polvec_0 \cdot \int \psi_f^*(\vb{r}) e^{i \vb{k}\cdot\vb{r}} \vb{\nabla}\psi_i(\vb{r}) d^3 r \bigg) \bigg(  \polvec^\prime_0 \cdot \int \psi_f(\vb{r}) e^{-i \vb{k}^\prime\cdot\vb{r}} \vb{\nabla}\psi_i^*(\vb{r}) d^3 r \bigg)
\end{equation}
Referring to Catena \textit{et al}~\cite{Catena:2019gfa}, we can then apply the definition of the vectorial form factor (eq B18, but with some changes made to keep the notation more consistent),
\begin{equation}
    \vb{f}_{1\to2}(\vb{q}) = \int d^3 r \psi^*_f (\vb{r}) e^{i \vb{q}\cdot\vb{r}} \frac{i \vb{\nabla}}{m_e} \psi_i (\vb{r}).
\end{equation}
Here the final state and initial state wave functions have quantum numbers $i = n,\ell,m$ and $f = p^\prime,\ell^\prime, m^\prime$ where $p^\prime$ is the final state electron momentum, and $\{n,\ell,m\},\{\ell^\prime,m^\prime\}$ are the initial and final quantum numbers, respectively. Applying this definition, we have
\begin{align}
    \Delta f^{\prime\prime} &= \dfrac{\pi \hbar^2}{m_e} \bigg( \polvec_0 \cdot (-i m_e) \vb{f}_{1\to2}(\vb{k})  \bigg) \bigg(  \polvec^\prime_0 \cdot (i m_e) \vb{f}^*_{1\to2}(\vb{k}^\prime) \bigg) \nonumber \\
    &=\pi \hbar^2 m_e \bigg(\polvec_0 \cdot \vb{f}_{1\to2}(\vb{k})\bigg) \bigg(\polvec_0^\prime \cdot \vb{f}^*_{1\to2}(\vb{k}^\prime)\bigg)
\end{align}
We can rewrite this expression with vector indices $i,j$ and take an average over polarization vectors using $\sum \epsilon_i \epsilon_j = \delta_{ij}$;
\begin{align}
    \Delta f^{\prime\prime} &= \pi \hbar^2 m_e (\polvec_0)_{i} (\vb{f}_{1\to2}(\vb{k}))_{i} (\polvec_0^\prime)_j (\vb{f}^*_{1\to2}(\vb{k}^\prime))_j \nonumber \\
    &= \pi \hbar^2 m_e \vb{f}_{1\to2}(\vb{k}) \cdot \vb{f}^*_{1\to2}(\vb{k}^\prime)
\end{align}
Then, in the Bragg limit $|\vec{k}| = |\vec{k}^\prime|$ and for spherically symmetric wavefunctions we may take
\begin{equation}
    \Delta f^{\prime\prime}(k) =\pi \hbar^2 m_e |\vb{f}_{1\to2}(k)|^2
\end{equation}
The incoherent form factors were also calculated and reported by Freeman~\cite{Freeman:a02679}.

Now we can express the Borrmann parameter with the quadrupole component of the imaginary form factor~\cite{wagenfield1986,PhysRev.144.216,persson_efimov};
\begin{equation}
    \epsilon \equiv D \bigg(1 - 2 \sin^2\theta_B \frac{\Delta f^{\prime\prime}_Q}{\Delta f^{\prime\prime}} \bigg) \frac{|S(h,k,l)|}{|S(0,0,0)|}
\end{equation}
where $D$ is the Debye-Waller factor accounting for thermal vibrations in anomalous absorption, $D = e^{-B s^2}$ where $s = \sin\theta / \lambda$ and $B$ is a temperature-dependent constant. The Debye-Waller factors for cryogenic temperatures can be found in ref.~\cite{Peng96} as well as fits to $\Delta f^{\prime\prime}$ for several pure materials of interest.

\begin{figure}
    \centering
    \includegraphics[width=0.6\textwidth]{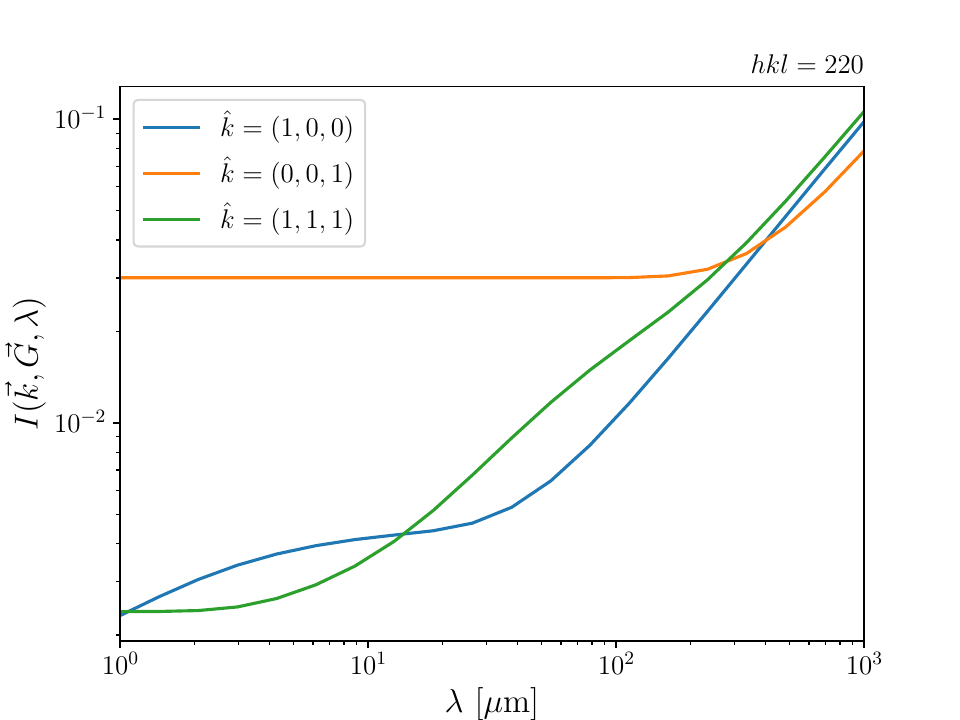}
    \caption{The absorption factor $I(\vec{k}, \vec{G}, \lambda)$ as a function of the mean free path $\lambda = 1/\mu$ for a crystal of cubic volume with side length 5 cm.}
    \label{fig:abs_factor}
\end{figure}

\begin{figure}[h]
    \centering
    \includegraphics[width=0.6\textwidth]{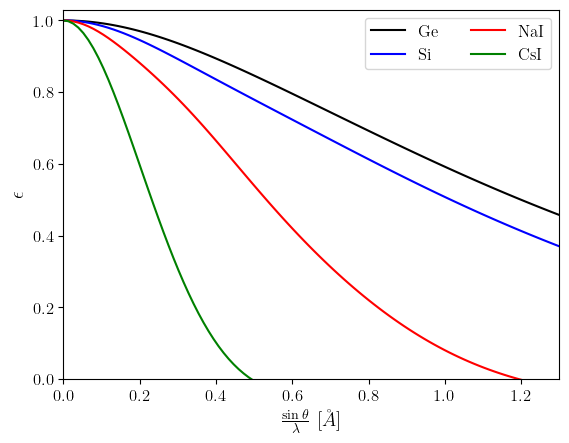}
    \caption{Borrmann parameter $\epsilon$ as a function of the momentum transfer $\sin\theta / \lambda = |\vec{G}|/4\pi$ for several crystal materials.}
    \label{fig:borrmann_by_mat}
\end{figure}

The absorptive part of the coherent sum that remains after the Laue condition is met is
\begin{equation}
   I(\vec{k},\vec{G},\lambda) \equiv \sum_{j\neq i}^N \sum_{i=1}^N e^{-\frac{(\vec{k}-\vec{G})}{|\vec{k}-\vec{G}|}\cdot(\vec{r}_i - \vec{r}_j))/(2\lambda)}
\end{equation}
Taking the Ge lattice as an example, with lattice constant $d = 5.657$  \AA, we evaluate $I(\vec{k},\vec{G},\lambda)$ numerically by constructing a lattice of $N$ Ge atoms. Since computing the full sum for a real crystal of centimeter length scale would require a huge number of evaluations $(\propto N^2)$, we take a sparse sampling of $N$ atoms across the physical crystal volume such that the sum is computationally feasible. The sum can then be evaluated in increments of increasing $N$ to test for convergence. We find that a lattice of around $N\simeq 10^4$ atoms in a cubic geoometry is enough to obtain a convergent error of around 5\%. Some evaluations of $I(\vec{k},\vec{G},\lambda)$ as a function of varying mean free path $\lambda$ are shown in Fig.~\ref{fig:abs_factor} for several choices of scattering planes $\vec{G}$ and incoming wavevectors $\vec{k}$.

One interesting phenomenon that can be seen in Fig.~\ref{fig:abs_factor} is that there are certain choices of $\vec{k}^\prime = \vec{k} - \vec{G}$ such that $\vec{k}^\prime \cdot (\vec{r}_i - \vec{r}_j) = 0$. In this special circumstance, while many of the terms in the coherent sum will tend to zero with decreasing $\lambda$, the terms where this dot product is zero will survive. What this means physically is that the plane in which $\vec{r}_i - \vec{r}_j$ lies will avoid the decoherence from absorption as long as it remains orthogonal to $\vec{k}^\prime$. This relation can be made more apparent by considering the dot product under the Bragg condition;
\begin{align}
    \hat{k}^\prime \cdot (\vec{r}_i - \vec{r}_j) &= \bigg(\frac{\vec{G}}{2 \vec{k}\cdot\hat{G}} - \frac{\vec{G}}{k}\bigg)\cdot(\vec{r}_i - \vec{r}_j) = 0
\end{align}
where we take $\hat{k} = (\cos\phi\sin\theta, \sin\phi\sin\theta,\cos\theta)$, solving this equation for $\theta$ in the $hkl = 400$ case gives
\begin{equation}
   \theta = \cot ^{-1}\left(\frac{n_x \cos (\phi )-n_y \sin (\phi )}{n_z}\right)+\pi  c_1
\end{equation}
for $n_x, n_y, n_z, c_1 \in \mathbb{Z}$. This defines a family of lattice points that remain in the absorption sum $I$ even in the limit $\lambda \to 0$, resulting a lower bound on $I$ as shown for some example choices of $\hat{k}$ in Fig.~\ref{fig:abs_factor}. This effect is similar in nature to the Laue-case diffraction enhancements where the photoconversion occurs down the scattering planes, minimizing the absorption, as studied in ref.~\cite{Yamaji:2017pep}.

\begin{figure}
    \centering
    \includegraphics[width=0.43\textwidth]{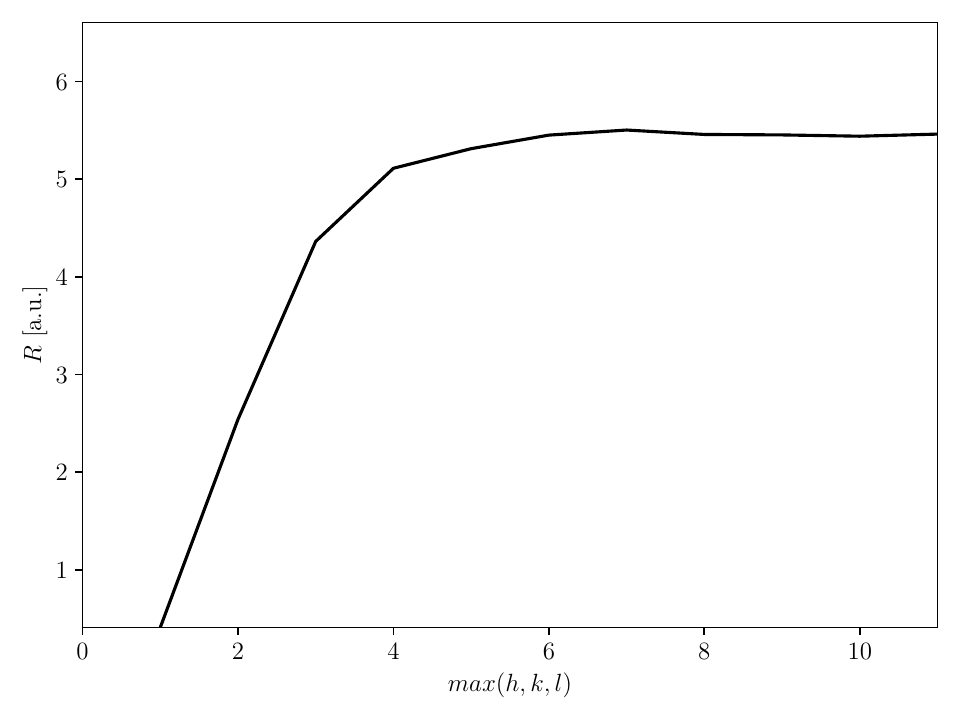}
    \includegraphics[width=0.47\textwidth]{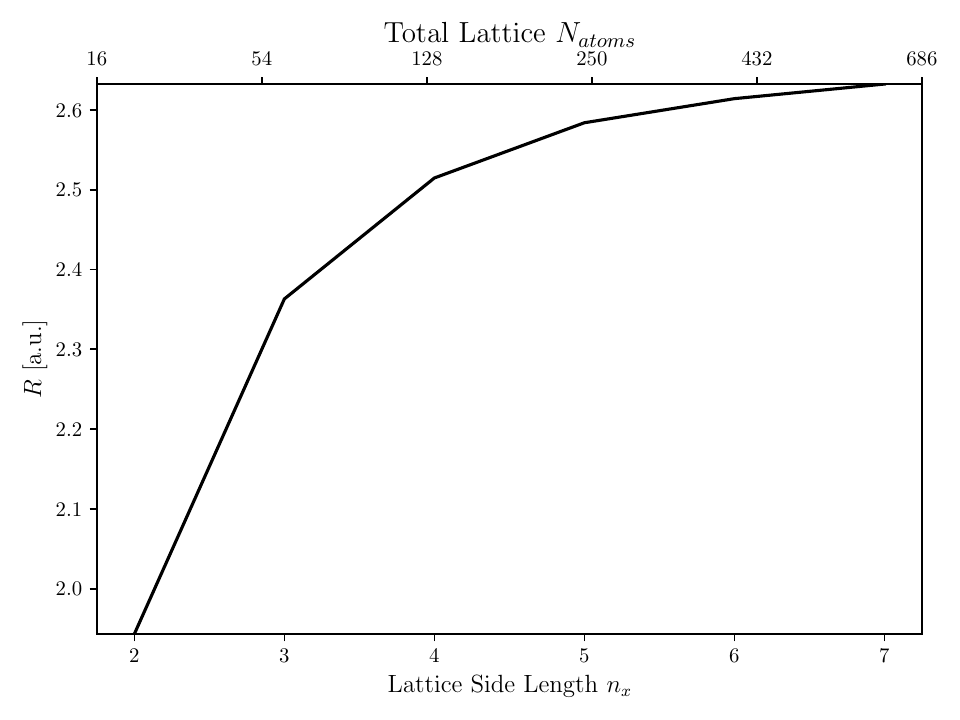}
    \caption{Left: convergence rate for the sum over $\vec{G}$ in the event rate $R$ as a function of the cutoff $max(h,k,l)$. Right: convergence of the event rate with the absorption sum $I(\vec{k},\vec{G})$ for a given simulated lattice size.}
    \label{fig:convergence_G}
\end{figure}

We employ a numerical cutoff on the sum over reciprocal lattice planes $\vec{G}(hkl)$ at $max(h,k,l)=6$, where a test for convergence of the event rate shows that the corrections beyond add corrections less than 5\% (see Fig.~\ref{fig:convergence_G}). Additionally, the values of the structure factor for a Ge diamond cubic lattice are tabulated in Table~\ref{tab:miller_table}, ordered by decreasing $|S(hkl)|^2/|\vec{G}|^2$ as they appear in the event rate sum.

\begin{figure}[h]
    \centering
    \includegraphics[width=0.8\textwidth]{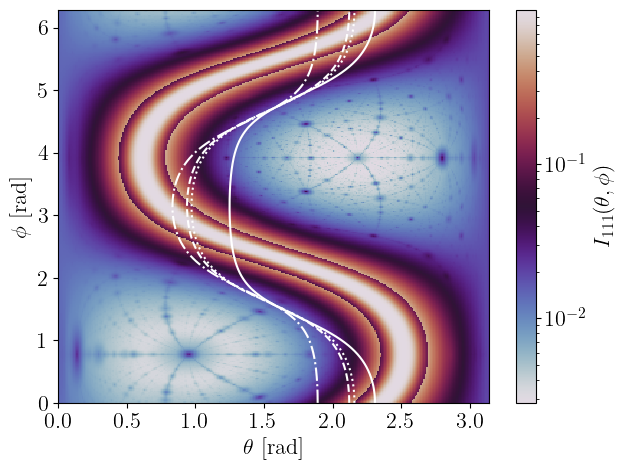}
    \caption{The absorption factor $I(\vec{k}, \vec{G})$ for the $hkl=111$ plane as a function of the incoming ALP direction $\hat{k}(\theta,\phi)$ when the bragg condition is satisfied and the mean free path is given by the Borrmann anomalous absorption coefficient. Here the we take a crystal of cubic volume with side length 5 cm.  The solid white lines trace sample paths of the daily solar angle in January (solid), March (dashed), June (dash-dotted), and September (dotted) at the latitude and longitude of the Gran Sasso site.}
    \label{fig:abs_factor2d}
\end{figure}

In Fig.~\ref{fig:abs_factor2d} the absorption factor $I$ is shown for the plane $\vec{G}(1,1,1)$ as a function of azimuthal and polar angles of the incoming axion momentum $\theta, \phi$ under the Bragg condition. This fixes $k = E_\gamma$ for a given $(\theta, \phi)$, and therefore the attenuation length $\lambda$ given by Eq.~\ref{eq:anomalous_depth}. We see a two prominent features of mitigated absorption in the $S$-shaped band (tracing out a great circle on the 2-sphere), where (i) $I\to 1$ as these $(\theta,\phi)$ combinations correspond to larger energies where the photon absorption cross section falls off as we move further into the $S$, and (ii) there is a jump discontinuity in the $S$-band due to an absorption edge in the photoelectric cross section for germanium at around 11 keV.

\makeatletter
\newcommand{\asteriskfootnote}[1]{%
  \begingroup
  \renewcommand{\thefootnote}{}
  \footnotetext{#1}%
  \endgroup
}
\makeatother

\newcommand\blfootnote[1]{%
  \begingroup
  \renewcommand\thefootnote{\huge\textasteriskcentered}\footnote{#1}%
  \addtocounter{footnote}{-1}%
  \endgroup
}

\newcommand{\chapterWithAsterisk}[1]{%
  \chapter[#1]{#1\,\huge\textasteriskcentered\asteriskfootnote{*\,Parts
of this chapter are adapted from ref.~\cite{Dent:2020jhf} with permission from Adrian Thompson and co-authors.}}
}

\chapterWithAsterisk{\uppercase{Laboratory Searches for Solar Axions at Direct Detection Experiments}}\label{ch:solar}

\section{Dark Matter Direct Detection Experiments}
Dark matter direct detection experiments, initially designed to search for WIMP-like dark matter, have been adapted more broadly as detectors of Beyond Standard Model (BSM) physics. Notable among the wide class of BSM physics searches at direct detection facilities is the extraordinary sensitivity to possible axion or axion-like particles coupling to Standard Model particles (SM) \cite{Akerib:2017uem,Fu:2017lfc,Abe:2018owy,Armengaud:2018cuy,Aprile:2019xxb,Wang:2019wwo,Aralis:2019nfa}. By examining electronic recoils produced by a solar axion flux through the detector, these searches have probed a variety of $a$-SM couplings including axion-electron, axion-photon, and axion-nucleon interactions.

In ref.~\cite{Dent:2020jhf}, we investigated inverse Primakoff scattering as a new detection channel at liquid xenon based direct detection experiments. We showed that sole use of the coupling $g_{a\gamma}$ can fit the recent XENON1T excess of electron recoils in their low energy (1-30 keV) data, with a rise above the background-only model occurring below 7~keV~\cite{Aprile:2020tmw}. Although a null result was later found with 1.16 tonne-years of exposure~\cite{XENON:2022ltv}, we found at the time that the fitting of the excess is free of the leading helioscope CAST constraint for $m_a\gtrsim 0.03$ eV. However, even then the tension associated with the astrophysical constraints, HB star cooling from $R$ parameter measurements, was still 8$\sigma$~\cite{DiLuzio:2020jjp}.

\begin{figure}
    \centering
    \includegraphics[width=0.8\textwidth]{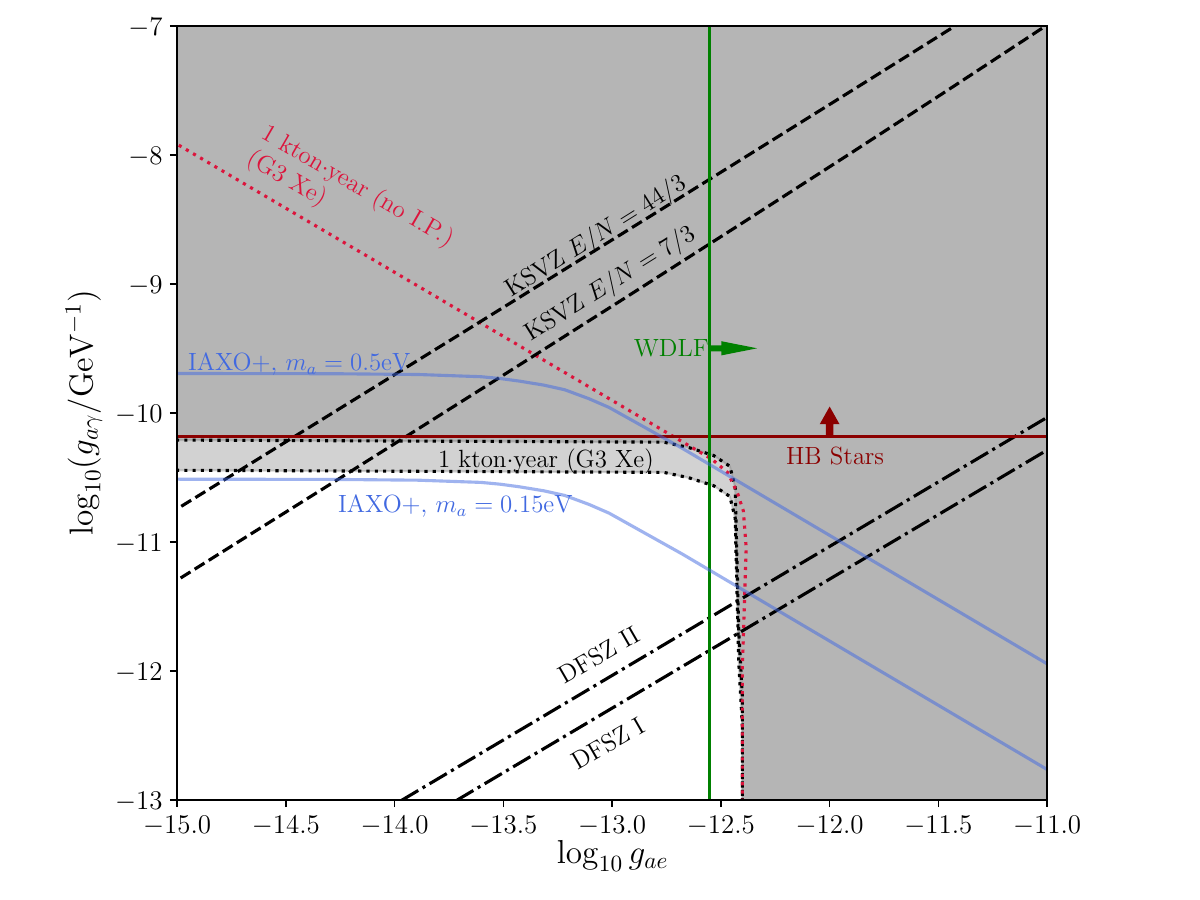}
    \caption{Projections of the 95\% CL future exclusions (using the Wigner-Seitz FF in light gray and RHF form factor in dark gray{We have changed this value of the screening radius from 2.45~\AA ~\cite{Walters_2006}\, to 0.685~\AA \, in accordance with the more common expression for the Wigner-Seitz radius~\cite{Ashcroft}, which accounts for at most a factor of $10$ reduction in the cross section at 1 keV down to a factor of 3 reduction at 5 keV with respect to the old value.}) set by G3 Xe over a 1 kton$\cdot$year exposure given background-only observations. The exclusion line for 1 kton$\cdot$year without inverse Primakoff (I.P.) scattering is shown for comparison (dotted red). We also show the IAXO+ projection (blue) which begins to lose sensitivity for $m_a \gtrsim 0.01$ eV. Reprinted from ref.~\cite{Dent:2020jhf} with permission from Adrian Thompson and co-authors.}
    \label{fig:xenon1t_fit_gae_gagamma}
\end{figure}

The more interesting prospect that emerged out of this analysis was that next-generation xenon experiments were projected to overcome the HB stars limit, and for $g_{ae}=10^{-13}$, the 2.4$\sigma$ hint region of stellar cooling can be probed within $1\sigma$; see Fig.~\ref{fig:xenon1t_fit_gae_gagamma}. In addition, these future bounds would be applicable for masses $m_a < 1$ keV, covering complementary regions of parameter space (including that of KSVZ axions) for which future helioscopes, such as IAXO, start to lose sensitivity near $m_a \gtrsim 0.01$ eV. A similar region of the $g_{a\gamma}-m_a$ space will also be investigated at LZ~\cite{Akerib:2019fml} and SuperCDMS SNOLAB~\cite{Agnese:2016cpb}, where the reach for $g_{a\gamma}$ needs to be scaled for the new detector type roughly by  $\sqrt{M_{D} Z^2_{D}/M_{Xe} Z^2_{Xe}}$ (where $M_D$ is the detector mass and $Z_D$ is the atomic number of the detector nucleus) for the same exposure.

\begin{figure}
    \centering
    \includegraphics[width=0.6\textwidth]{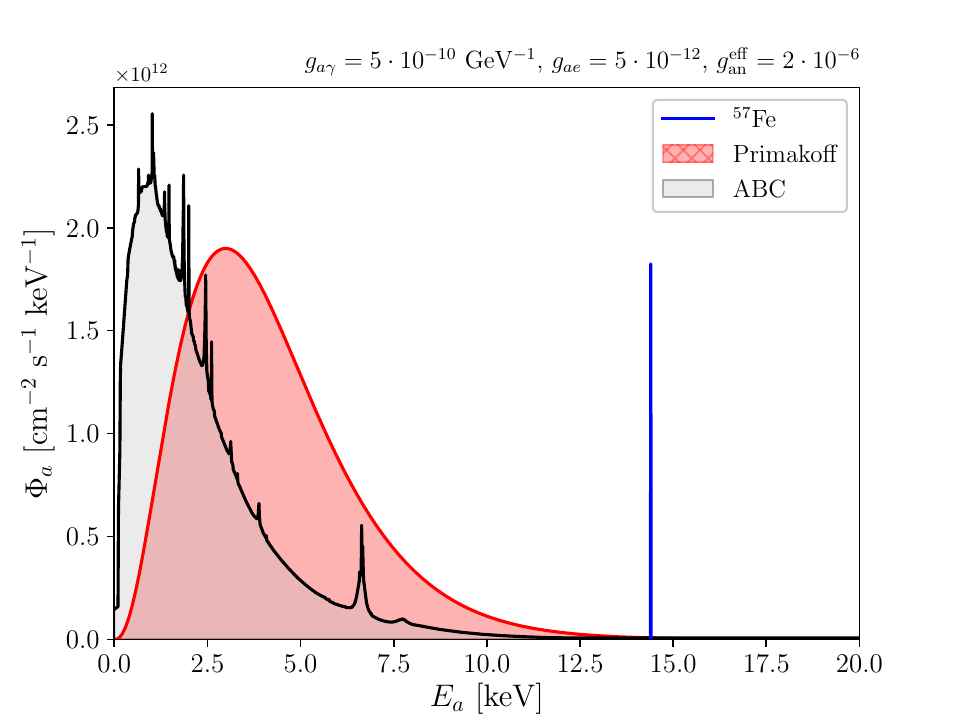}
    \caption{Solar axion fluxes at the Earth's surface are shown for the ABC, Primakoff, and $^{57}$Fe components. The bulk shape of the ABC component is due to axion-bremsstrahlung and Compton scattering, $e^- I \to e^- I\, a$ and $e^- \gamma \to e^- a$, while the numerous peaks are due to atomic transitions in $I^* \to I a$ and $e^- I \to I^- a$. The Primakoff flux exhibits a smooth thermal distribution from the coherent conversion of photons into axions, while the $^{57}$Fe flux is monoenergetic and is expected to broaden out from detector energy response effects. Reprinted from ref.~\cite{Dent:2020jhf} with permission from Adrian Thompson and co-authors.}
    \label{fig:solar_alp_flux}
\end{figure}

There are three prominant sources of solar axion flux that we considered with keV-scale energy spectra, each with a dependence on a different axion coupling parameter. First is the the ``ABC" flux, driving axion production from \textbf{A}tomic de-excitation and recombination, \textbf{B}remmstrahlung, and \textbf{C}ompton scattering processes~\cite{Redondo:2013wwa}, dependent on the $g_{ae}$ coupling. Next, the Primakoff production process, $\gamma Z \to a Z$, occurs via the $g_{a\gamma}$ coupling through the $t$-channel exchange of a virtual photon scattering with electrons or ions in the solar interior~\cite{Andriamonje:2007ew,Hagmann:2008zz}. Finally, de-excitation of $^{57}$Fe in the sun can produce a monoenergetic axion population at 14.4~keV~\cite{Moriyama:1995bz,Andriamonje:2009dx,Alessandria:2012mt}. This flux would arise from an effective axion-nucleon coupling $g_{\rm an}^{\rm eff} = -1.19g_{\rm an}^0 + g_{\rm an}^3$, where $g_{\rm an}^{0(3)}$ are the isoscalar (isovector) coupling constants for the nucleons \cite{Haxton:1991pu,Alessandria:2012mt}. Each of these flux components are shown in Fig.~\ref{fig:solar_alp_flux}.

We considered inverse Primakoff scattering in addition to the axioelectric absorption process outlined in the analysis performed by XENON1T (see also~\cite{Dimopoulos:1985tm,Ljubicic:2004gt,PhysRevD.35.2752,Alessandria:2012mt}). Also, it is possible that axions undergo inverse Compton scattering off electrons at rest in LXe, $a e^- \rightarrow \gamma e^-$~\cite{Avignone:1988bv}, but this is a subdominant process ($\propto Z$) in comparison to axioelectric scattering ($\propto Z^5$). If both axion-photon and axion-electron couplings are present, there are interference terms present in the total matrix element of the combined processes, which are also subdominant, but we included them as a matter of completeness.

\begin{figure}
    \centering
    \includegraphics[width=0.6\textwidth]{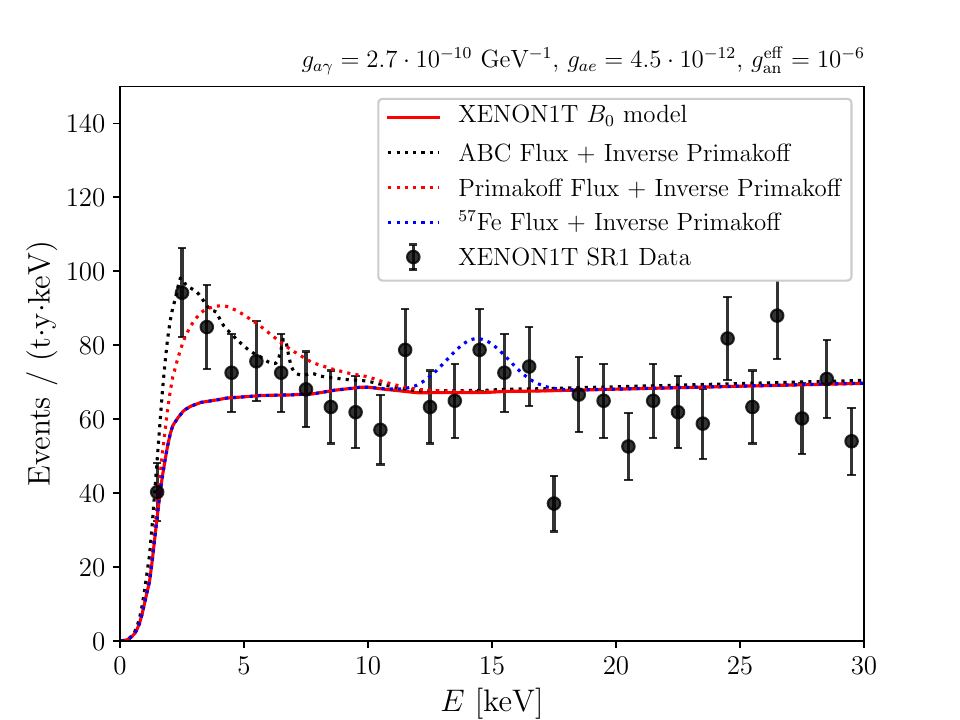}
    \caption{The event rate distributions for inverse Primakoff scattering in LXe for a tonne-year exposure from Primakoff-produced, ABC-produced, and $^{57}$Fe-produced axions are shown for select choices of axion couplings, and added to the ``$B_0$" background model. Reprinted from ref.~\cite{Dent:2020jhf} with permission from Adrian Thompson and co-authors.}
    \label{fig:xenon1t_spectra}
\end{figure}

To predict the event spectra from axions produced through ABC, Primakoff, and $^{57}$Fe, we convolved the fluxes in each case with the total cross sections, for inverse Primakoff scattering or axioelectric absorption, and multiply by the detector efficiency~\cite{Aprile:2020tmw}. In addition, we approximated the detector response for the energy resolution by convolving the simulated differential event distribution with an energy-dependent Gaussian smearing function~\cite{Aprile:2020tmw,Aprile:2020yad,Aprile:2019dme}. The event distribution for Primakoff-produced axions that undergo inverse Primakoff scattering in the LXe fiducial volume over a tonne-year exposure is shown in Figure~\ref{fig:xenon1t_spectra}.

In Fig.~\ref{fig:xenon1t_fit_gae_gagamma} (bottom), I show the next-generation xenon (G3 Xe) constraint (with a 1 kilotonne-year exposure~\cite{Szydagis:2016few}) where we found that the 95\% CL can overcome even the HB stars constraint and start exploring the mild hint (2.4$\sigma$) region of stellar cooling within 1$\sigma$. Interestingly, this is only possible with the inclusion of the inverse Primakoff channel since without this channel the constraint could be worse by a few orders of magnitude. We also found that our projected sensitivity for a 1 kton$\cdot$year exposure at a G3 LXe experiment is competitive with future helioscope experiments. The proposed DARWIN detector would achieve a 200 tonme-year exposure~\cite{Aalbers:2016jon}, thereby covering the current HB Stars constraint. Comparing the 1 kt$\cdot$year projection against the projected sensitivities for IAXO+ with masses $m_a > 0.1$ eV, one can see where the sensitivity begins to diminish for larger masses~\cite{Armengaud:2019uso} and the direct detection experiments play an important role as a broadband search over ALP masses. Additionally, future direct detection experiments with directional sensitivity would be able to use the directional information to reduce backgrounds and further increase their sensitivity to solar axions. This is especially useful in the Primakoff channel, where the axion's incoming direction is approximately preserved due to the low momentum transfer / forward scattering by the photon in the relativistic limit.

\section{Pushing the Sensitivity Envelope with Bragg-Primakoff Conversion}

Using the event rate formula for Bragg-Primakoff coherent scattering with a perfect crystal worked out in \S~\ref{ch:crystals}, the event rate in an energy window $[E_1, E_2]$ is
\begin{align}
\label{eq:event_rate_full_coherency}
    \dfrac{dN}{dt} = \pi g_{a\gamma}^2 (\hbar c)^3 \dfrac{V}{v_\textrm{cell}^2} &\sum_{\vec{G}} I(\vec{k},\vec{G}) \bigg[ \dfrac{d\Phi_a}{dE_a}  |F_j (\vec{G}) S_j(\vec{G}) |^2\frac{4(\hat{G}\cdot\hat{k})^2 (1 - (\hat{G}\cdot\hat{k})^2)}{|\vec{G}|^2} \mathcal{W} \bigg]
\end{align}
where $S_j$ is the crystal structure factor (see Ch.~\ref{ch:crystals}), $F_j$ is the atomic form factor for species $j$, and $d\Phi_a / dE_a$ is the solar axion flux from Primakoff scattering and photon coalescence in the sun~\cite{Raffelt:1996wa, PhysRevD.18.1829, DiLella:2000dn}.
The sum over the reciprocal lattice vectors $\vec{G}$ effectively counts the contributions to the coherent scattering from each set of lattice planes, illustrated in Fig.~\ref{fig:csi_planes}.

\begin{figure}
    \centering
    \includegraphics[width=0.8\textwidth]{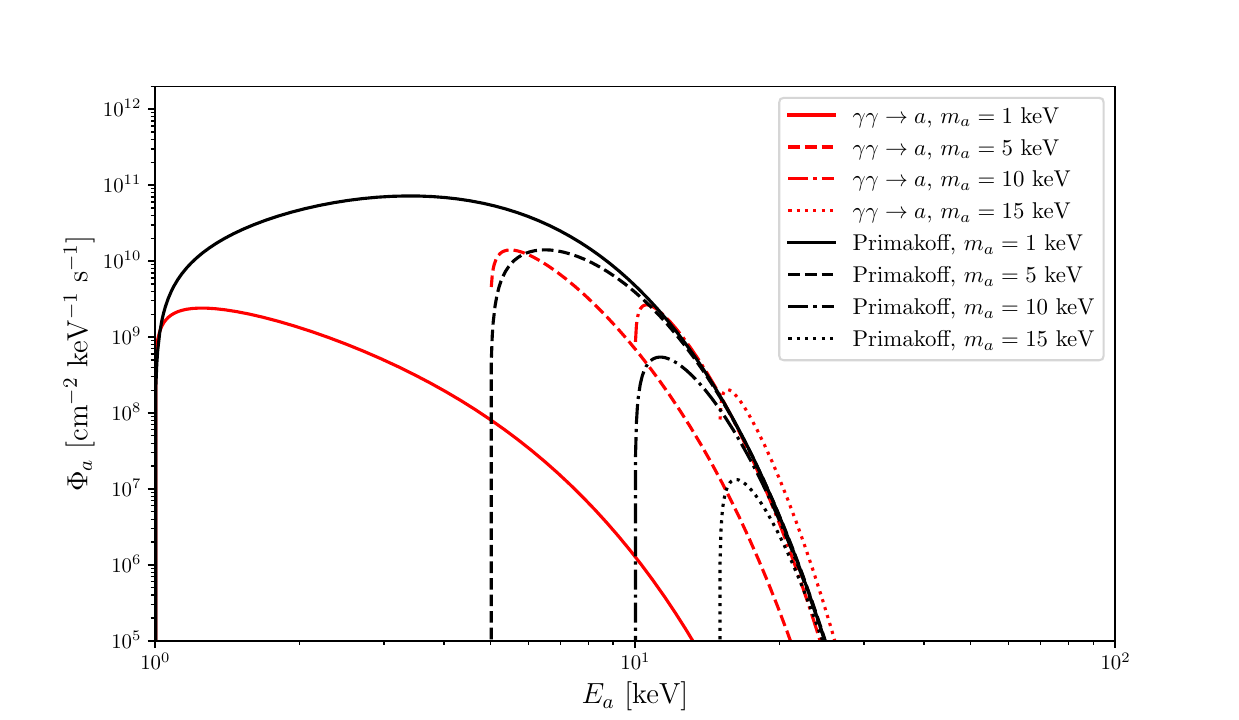}
    \caption{Solar axion fluxes from Primakoff conversion of photons and photon coalescence are shown for several mass benchmarks.}
    \label{fig:solar_fluxes}
\end{figure}

For this analysis we may be also interested in axion masses that are heavier, around $\sim$1-10 keV. For this we use the parameterization appearing in ref.~\cite{DiLella:2000dn} for massive axion production in the sun; the flux parameterizations are repeated here for convenience
\begin{align}
\label{eq:solar_fluxes}
    \dfrac{d\Phi_{\gamma\to a}}{dE_a} &= \dfrac{4.20\cdot 10^{10}}{\textrm{cm}^{-2}\textrm{s}^{-1}\textrm{keV}^{-1}} \bigg(\dfrac{g_{a\gamma}}{10^{-10}\textrm{GeV}^{-1}}\bigg)^2
  \dfrac{E_a p_a^2}{e^{E_a/1.1} -0.7} (1 + 0.02 m_a)\\
    \dfrac{d\Phi_{\gamma\gamma\to a}}{dE_a} &= \dfrac{1.68\cdot 10^9}{\textrm{cm}^{-2}\textrm{s}^{-1}\textrm{keV}^{-1}} \bigg(\dfrac{g_{a\gamma}}{10^{-10}\textrm{GeV}^{-1}}\bigg)^2 m_a^4 p_a 
    \bigg(1 + 0.0006 E_a^3 + \frac{10}{E_a^2 + 0.2}\bigg) e^{-E_a}
\end{align}
where $\Phi_{\gamma \to a}$ is the Primakoff solar flux and $\Phi_{\gamma\gamma\to a}$ is the flux resulting from resonant photon coalescence, both in units of $\textrm{cm}^{-2}\textrm{s}^{-1}\textrm{keV}^{-1}$, given for axion energy and momentum $E_a$ and $p_a$ in keV, and for the coupling $g_{a\gamma}$ in GeV$^{-1}$. The solar axion flux from photon coalescence and Primakoff conversion is shown in Fig.~\ref{fig:solar_fluxes}.

The time dependence is encoded in the solar position, which we can express through $\hat{k} = (\cos\phi \sin\theta, \sin\phi \sin\theta, \cos\theta)$ for $\theta = \theta(t)$ and $\phi = \phi(t)$. For the solar angle as a function of time and geolocation, we use the NREL solar position algorithm~\cite{nrel}. 

\begin{figure*}[th]
    \centering
    \includegraphics[width=0.45\textwidth]{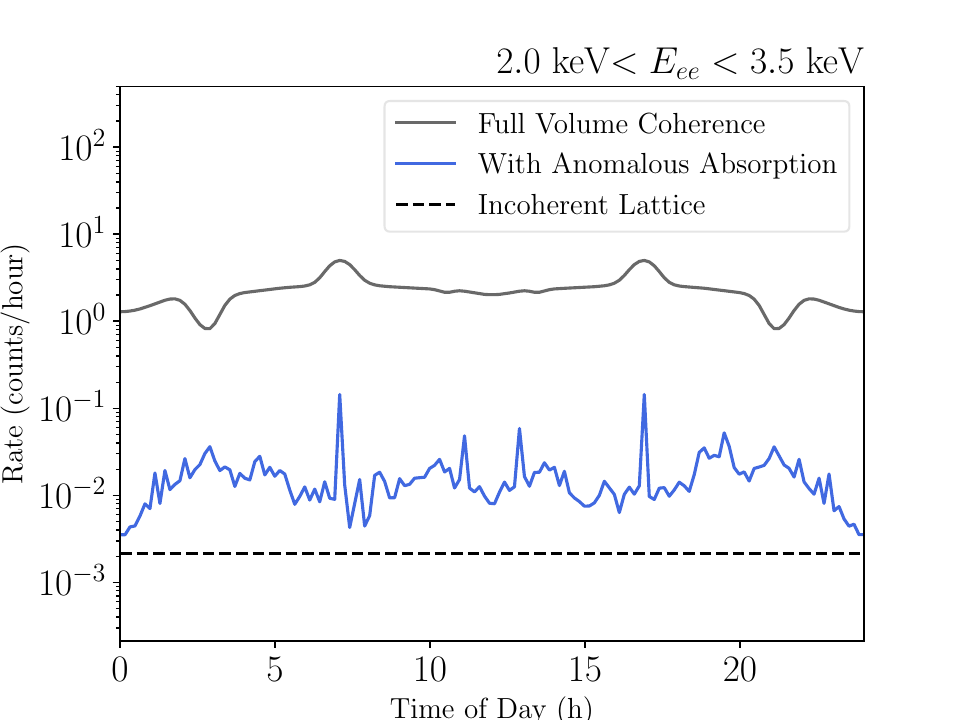}
    \includegraphics[width=0.45\textwidth]{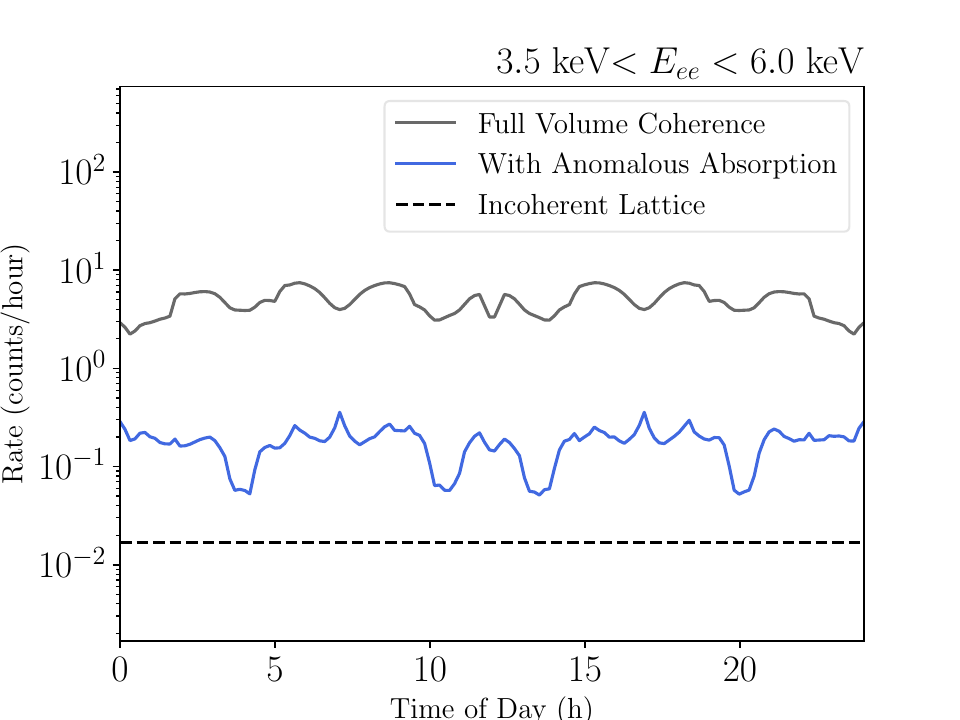}\\
    \includegraphics[width=0.45\textwidth]{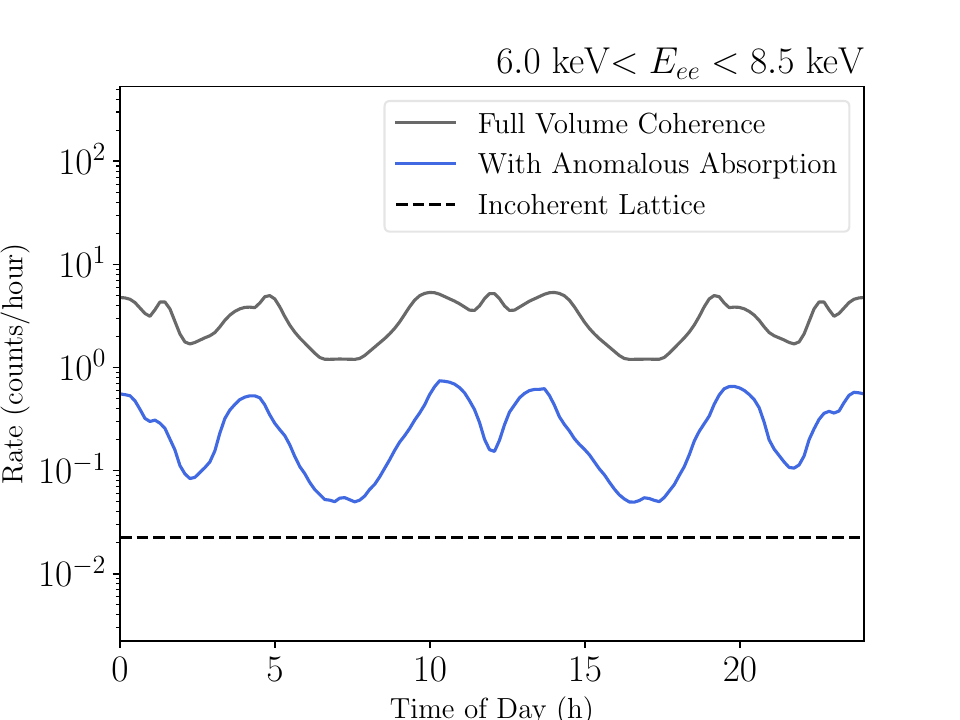}
    \includegraphics[width=0.45\textwidth]{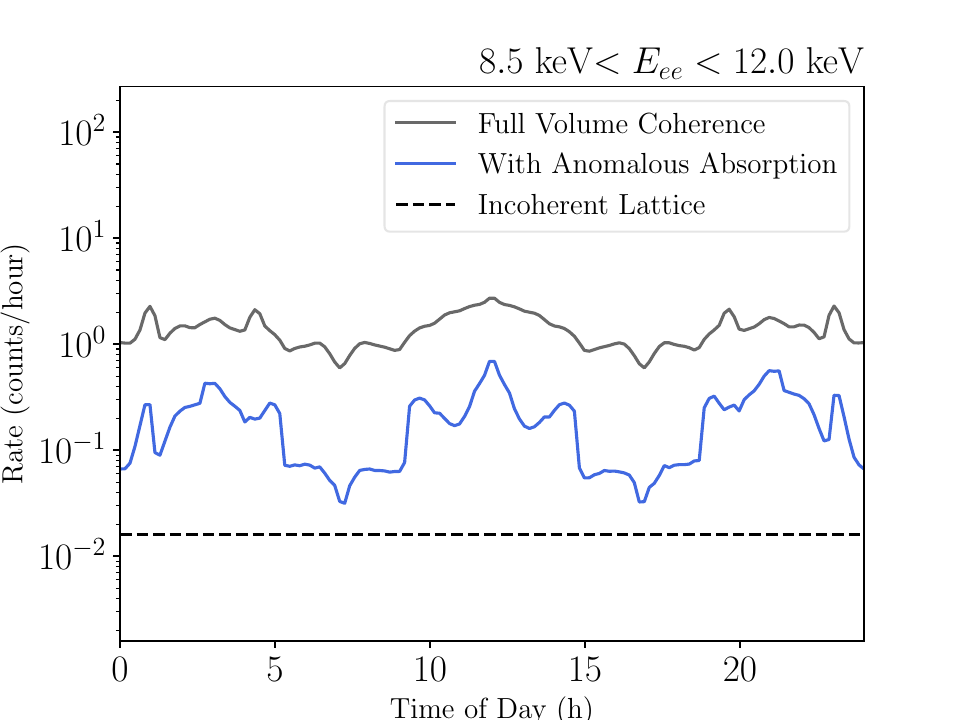}
    \caption{Solar ALP scattering rates in a 250 cm$^3$ Ge crystal, comparing the rates with full volume coherency to ours with anomalous absorption effects included through the absorption factor $I(\vec{k},\vec{G})$. Here we fix the coupling $g_{a\gamma} = 10^{-8}$ GeV$^{-1}$, energy resolution to be $\Delta = 1.0$ keV (for $E_{ee} < 6$ keV) and $\Delta = 1.5$ keV (for $E_{ee} > 6$ keV).}
    \label{fig:ge_rates}
\end{figure*}

The corresponding event rates for various energy windows are shown in Fig.~\ref{fig:ge_rates} for Ge crystal, where we compare the relative enhancements with and without the Borrmann effect to the case of full-volume coherency and to the case of incoherent scattering on an amorphous lattice\footnote{Atomic Primakoff scattering is still coherent here; only the coherency at the level of the lattice is lost for the sake of comparison with scattering on amorphous materials.}. The fluctuating features in the event rate are the result of the sum over $\vec{G}$ which contributes to the Bragg peaks. Here we have assumed a volume of 260 cm$^3$ (corresponding roughly to the volumetric size of a SuperCDMS germanium module), and so the relative suppression for each $\vec{G}$ lattice plane goes like $V^{1/3} / \lambda(\vec{k},\vec{G})$, giving a suppression on the order of $10^2$ compared to the full-volume coherency assumption.

\begin{figure}[h]
    \centering
    \includegraphics[width=0.48\textwidth]{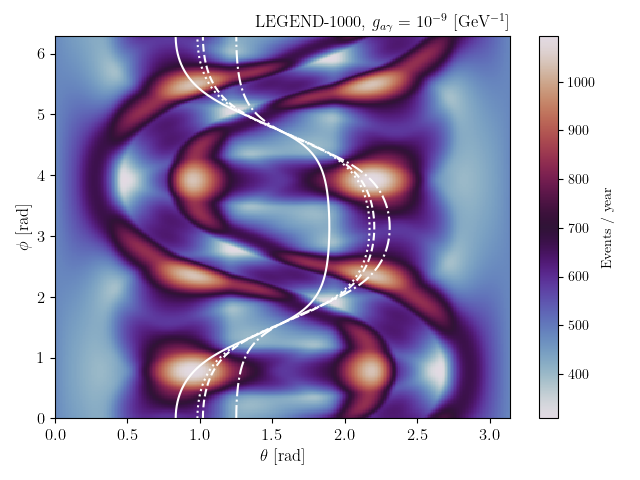}
    \includegraphics[width=0.48\textwidth]{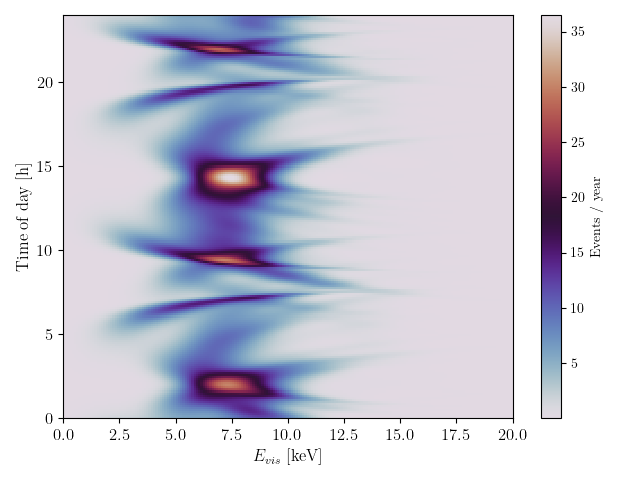}
    \caption{Right: Event rates in germanium as a function of incident angles $\theta, \phi$ for the integrated energy range $(1, 10)$ keV and for reciprocal lattice planes $(h,k,l)$ up to $max\{h,k,l\} = 5$. The solid white lines trace sample paths of the daily solar angle in January (solid), March (dashed), June (dash-dotted), and September (dotted) at the latitude and longitude of the Gran Sasso site. Left: Differential energy-time event rate with energy resolution $\Delta = 2.5$ keV. The time of year was taken to be January at the latitude and longitude of the Gran Sasso site.}
    \label{fig:2d_rates}
\end{figure}

The time-dependence can be visualized further by viewing the event rates as a function of incident angles integrated across the whole solar axion energy window, as shown in Fig.~\ref{fig:2d_rates} (left). Depending on the time of year, different sets of Bragg peaks will be traced over during the day, inducing a annual modulation in addition to the intra-day modulation of the signal.
Since the time of day fixes the solar zenith and azimuth $(\theta, \phi)$, we can finally show the spectrum of the Primakoff signal as a function of energy deposition and time of day; see Fig.~\ref{fig:2d_rates} (right).

\begin{figure}[h]
    \centering
    \includegraphics[width=0.9\textwidth]{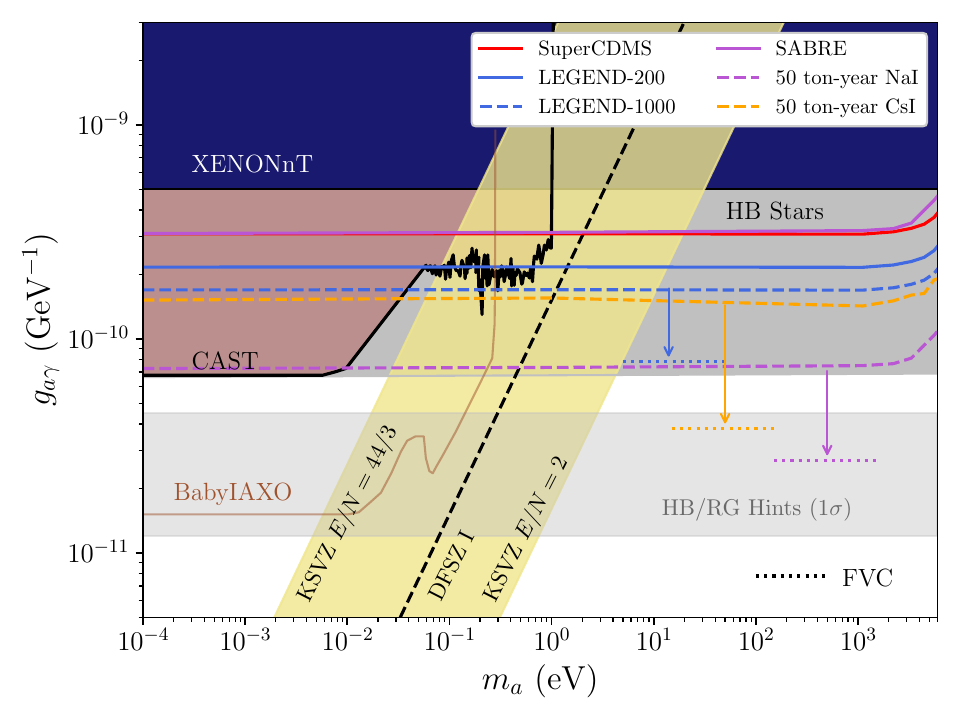}
    \caption{Sensitivity projections using the exposures listed in Table~\ref{tab:detectors} for SuperCDMS, LEGEND-200, LEGEND-1000, SABRE, and multi-tonne setups with absorption effects. The extension of these sensitivities to full volume coherency are indicated with the arrows.}
    \label{fig:sensitivity}
\end{figure}

We forecast the event rates for SuperCDMS~\cite{SuperCDMS:2022kse}, LEGEND-200, LEGEND-1000, SABRE, in addition to envisioned multi-ton setups, with detector specifications listed in Table~\ref{tab:detectors}. For the background-free limits, we look for the Poisson 90\% CL corresponding to $\simeq 3$ events observed for a given exposure. The limits on the $(g_{a\gamma} - m_a)$ parameter space are shown in Fig.~\ref{fig:sensitivity}, where we show projections assuming full volume coherency without absorption effects (top) as well as with absorption effects (bottom). In the case where we assume full volume coherency, we forecast the multi-kg-scale SuperCDMS setup to set the first laboratory based limits for $m_a > 1$ eV beyond the bounds set by XENONnT~\cite{XENON:2022mpc}. Ton-scale setups like LEGEND-200 and LEGEND-1000 can reach further, probing couplings up to the existing bounds fom HB Stars~\cite{Ayala:2014pea,Giannotti:2015kwo} and CAST~\cite{CAST:2017uph} for masses $m_a \lesssim 10$ keV, losing sensitivity for higher masses where the axion production rates from photon coalescence and Primakoff scattering diminish (see also Fig.~\ref{fig:solar_fluxes}). These reach more than an order of magnitude lower in the coupling than previous Bragg-Primakoff solar axion searches, also shown here for DAMA~\cite{Bernabei:2004fi}, CUORE~\cite{Li:2015tyq}, Edelweiss-II~\cite{Armengaud_2013}, SOLAX~\cite{PhysRevLett.81.5068}, COSME~\cite{COSME:2001jci}, CDMS~\cite{CDMS:2009fba}, and Majorana~\cite{Majorana:2022bse}. The QCD axion parameter space is shown here for the Kim-Shifman-Vainshtein-Zakharov (KSVZ) and Dine-Fischler-Srednicki-Zhitnitsky (DFSZ) type benchmark models, where the range is defined by taking the anomaly ratios of $E/N = 44/3$ to $E/N = 2$~\cite{DiLuzio:2020wdo}, although the space of heavier masses is also possible in high-quality axion models and other scenarios, e.g. refs.~\cite{Kivel:2022emq, Valenti:2022tsc}. To probe this model parameter space beyond the existing bounds from CAST and HB stars, multi-ton scale experiments are needed. 

However, with the effects of absorption included (Fig.~\ref{fig:sensitivity}, bottom), the suppression of the event rate brings our projections for LEGEND, SuperCDMS and SABRE to test parameter space already excluded by HB stars constraints. However, the LEGEND-1000 background-free scenario is projected to have the leading laboratory-based sensitivity beyond the existing limits from CAST and XENONnT, but multi-ton CsI and NaI setups would extend this to nearly cover the HB stars exclusion. The existing bounds from Edelweiss-II, COSME, SOLAX, CDMS, DAMA, and Majorana are not shown here, but their exclusions would necessarily shift to larger coupling values to account for absorption effects in the Bragg-Primakoff rates, depending on the volume and detector material. For the future projections, note that the relative reach between NaI and CsI crystals is relatively suppressed when absorption is included here, due to the behavior of the imaginary form factor for CsI giving only modest Borrmann enhancements at the lower reciprocal lattice planes; see Fig.~\ref{fig:borrmann_by_mat}. In order to push the sensitivity envelope beyond the current bounds by CAST and HB stars, even with multi-ton setups, the absorption effects need to be mitigated in order to recover the event rate sensitivity in full-volume coherence. Some possibilities are discussed in the following section.

\begin{table*}[ht!]
    \centering
    \begin{tabular}{|l|l|c|c|c|c|}
    \hline
         Experiment & \thead{Module Mass\\$\times$no. Modules} & \thead{Total Mass} & \thead{Energy\\Resolution} & Threshold & Exposure (tonne-years) \\
         \hline
         SuperCDMS (Ge) & 1.4 kg $\times$ 18 & 25.2 kg & 2.5 keV & 1 keV & 0.1 \\
         SuperCDMS (Si) & 0.6 kg $\times$ 6 & 3.6 kg & 2.5 keV & 1 keV & 0.0144 \\
         LEGEND-200 (Ge) & 2.6 kg $\times$ 75 & 195 kg & 2.5 keV & 1 keV & 0.78 \\
         LEGEND-1000 (Ge) & 2.6 kg $\times$ 400 & 1 ton & 2.5 keV & 1 keV & 4.16 \\
         SABRE (NaI) & 2 kg $\times$ 25 & 50 kg & 1 keV & 1 keV & 0.15\\
         Ton-scale NaI & 2 kg $\times$ 2500 & 5 ton & 1 keV & 1 keV & 50 \\
         Ton-scale CsI & 2 kg $\times$ 2500 & 5 ton & 1 keV & 1 keV & 50 \\
         \hline
    \end{tabular}
    \caption{Assumed detector parameters for the SuperCDMS~\cite{SuperCDMS:2022kse}, LEGEND~\cite{LEGEND:2021bnm}, and SABRE~\cite{SABRE:2018lfp} configurations, as well as those taken for the NaI and CsI multi-tonne benchmarks. Exposures are based on 4-5 years of active run time.}
    \label{tab:detectors}
\end{table*}

We forecast the event rates for SuperCDMS~\cite{SuperCDMS:2022kse}, LEGEND-200, LEGEND-1000, SABRE, in addition to envisioned multi-tonne setups, with detector specifications listed in Table~\ref{tab:detectors}. For the background-free limits, we look for the Poisson 90\% CL corresponding to $\simeq 3$ events observed for a given exposure. The projected reach over the $(g_{a\gamma} - m_a)$ parameter space for these detector benchmarks is shown in Fig.~\ref{fig:sensitivity}, where we show projections including the effects of absorption and the Borrmann enhancement to the absorption length, in addition to the projected limits assuming full volume coherence (FVC), i.e. $I(\vec{k},\vec{G})\to1$, indicated by the arrows and dotted lines. 

The QCD axion parameter space is shown (yellow band) for the Kim-Shifman-Vainshtein-Zakharov (KSVZ) and Dine-Fischler-Srednicki-Zhitnitsky (DFSZ) type benchmark models, where the range is defined by taking the anomaly ratios of $E/N = 44/3$ to $E/N = 2$~\cite{DiLuzio:2020wdo}, although the space of heavier masses is also possible in high-quality axion models and other scenarios, e.g. refs.~\cite{Kivel:2022emq, Valenti:2022tsc}. To probe this model parameter space beyond the existing bounds from CAST and HB stars, when full volume coherence is maintained, multi-tonne scale experiments are needed. 

With the effects of absorption included, we project SuperCDMS, LEGEND, and SABRE to test parameter space unexplored by laboratory-based probes beyond the CAST and XENONnT constraints for $m_a \gtrsim 1$ eV, but already excluded by HB stars constraints. However, multi-tonne CsI and NaI setups would extend this to nearly cover the HB stars exclusion. Similar reach could in principle be found when considering the joint parameter space of multiple ALP couplings to photons, electrons, and nucleons~\cite{Dent:2020jhf}. For instance, by considering the $^{57}$Fe solar axion flux, one could look for 14.4 keV energy signatures and their Bragg-Primakoff peaks, although the sensitivity would likely contend with astrophysics constraints as well~\cite{Hardy:2016kme}.

The existing bounds from DAMA~\cite{Bernabei:2004fi}, CUORE~\cite{Li:2015tyq}, Edelweiss-II~\cite{Armengaud_2013}, SOLAX~\cite{PhysRevLett.81.5068}, COSME~\cite{COSME:2001jci}, CDMS~\cite{CDMS:2009fba}, and Majorana~\cite{Majorana:2022bse} are not shown here, but their exclusions would necessarily shift to larger coupling values to account for absorption effects in the Bragg-Primakoff rates, depending on the detector volume and material. Please see appendix~\ref{app:fvc} for the projected sensitivity and existing constraints assuming full volume coherence. In fig.~\ref{fig:sensitivity}, note that the relative reach between NaI and CsI crystals is relatively suppressed when absorption is included here, due to the behavior of the imaginary form factor for CsI giving more modest Borrmann enhancements at the lower reciprocal lattice planes; see Fig.~\ref{fig:borrmann_by_mat}. In order to push the sensitivity envelope beyond the current bounds by CAST and HB stars, even with multi-tonne setups, the absorption effects need to be mitigated. Some possibilities are discussed in the next section.

There may be ways to recover the sensitivity initially projected in the case of full-volume coherence by mitigating the loss of coherence due to absorption. These are of course speculative routes. Some of these routes for future work are enumerated below;
\begin{enumerate}
    \item Since the attenuation of the coherent volume is direction-dependent, as shown in Fig.~\ref{fig:abs_factor2d}, one could imagine optimizing a detector geometry such that the size and orientation relative to the incoming flux of axions is ideal, maximizing use of the Laue-type scattering and Borrmann effect to minimize the absorption. This would require precise knowledge of the crystal purity and plane orientation obtained from X-ray measurements.
    \item Along a similar vein, since the effects of absorption are minimized when the detector scale $V^{1/3}$ becomes comparable to the photon mean free path $\lambda$, one could instead prefer to use smaller detector volumes but with a large total mass partitioned into many individual modules. As long as each module is optically insulated from the others, the loss of coherence due to absorption will be contained within each module and the suppression to the event rate can be mitigated.
    \item It might be possible to apply the principles in this work to radioisotope experiments like those proposed in ref.~\cite{Dent:2021jnf}, where a keV-scale nuclear transition line (e.g. the 14.4 keV line of $^{57}$Fe) could source ALPs through a coupling to nucleons. Subsequent detection by an array of crystals encasing the radioactive source searching for transition photons of known energy Primakoff-converting in the crystal would leave a missing energy signature in the detector. By looking for disappearing keV-scale transitions the signal rate would enjoy the coherent enhancement relative to the incoherent scattering considered in ref.~\cite{Dent:2021jnf}.
    \item A dedicated keV photon source that would impinge on a crystal detector could fire at a fixed angle of incidence such that the event rate enhancement from the Borrmann effect and Laue effects are optimized and full volume coherence is restored as best as possible. One might achieve this with a keV laser~\cite{lanlLaser} or synchrotron sources in a similar fashion to LSW experiments~\cite{Yamaji:2018ufo, PhysRevLett.105.250405, PhysRevLett.118.071803, PhysRevD.105.035031}. By performing a similar ``missing'' photon search as the one discussed above, the event rate for the detection of missing energy will be proportional to $g_{a\gamma}^2$, rather than $g_{a\gamma}^4$ as in solar axion searches, greatly enhancing the sensitivity.\\
\end{enumerate}

In the case where we assume full volume coherence, shown in Fig.~\ref{fig:sensitivity}, dotted lines, ton-scale setups like LEGEND-200 and LEGEND-1000 can reach significantly smaller couplings, probing values of $g_{a\gamma}$ beyond the existing bounds fom HB Stars~\cite{Ayala:2014pea,Giannotti:2015kwo} and CAST~\cite{CAST:2017uph} for masses $m_a \lesssim 10$ keV, losing sensitivity for higher masses for which the axion production rates from photon coalescence and Primakoff scattering are diminished (see also Fig.~\ref{fig:solar_fluxes}). These reach more than an order of magnitude lower in the coupling than previous Bragg-Primakoff solar axion searches. 

Lastly, I will make some comments about extending this work to other important phenomenological probes. First, one can additionally consider non-relativistic ALPs, which may come from the dark matter halo, or from stellar basins like those considered in ref.~\cite{VanTilburg:2020jvl,DeRocco:2022jyq}. In these cases one can think about a modified Bragg-Primakoff condition, or instead a photoionization-like process which can actually dominate the detection channels in the non-relativistic regime~\cite{Wu:2022yei}. For this second option, we investigated this possibility in an alternative method to ref.~\cite{Wu:2022yei} that again makes use of the \texttt{DarkARC} code to compute the ionization amplitudes in inverse Primakoff photoionization; see appendix~\ref{app:pe_prim}. Secondly, the anomalies observed in white dwarf (WD), red giant (RG), and horizontal branch (HB) cooling data, known as the WD stellar cooling hints, have shown to be explained by ALP-assisted stellar cooling that would correspond to parameter space near the $g_{a\gamma}\lesssim 10^{-11}$ GeV$^{-1}$ level and below, and for non-vanishing $g_{ae} \simeq 10^{-13}$~\cite{Giannotti:2015kwo,Hoof:2018ieb,Ayala:2014pea}. These are also shown in Fig.~\ref{fig:sensitivity} by the gray band for the RG and HB cooling hints explained at the $1\sigma$ level in $g_{a\gamma}$, which extends down at the $2\sigma$ level to vanishing photon coupling, but can be explained at the 2$\sigma$ level by a non-vanishing electron coupling. Testing the ALP explanation is a further motivator of probing this parameter space, and an analysis that combines both electron and photon couplings in this context for future crystal detectors would be able to directly test this anomaly.

%
%
%
%


\chapter{\uppercase{Conclusion}}\label{ch:conclusion}

In this work, we have taken into account a more proper estimate of the effects of anomalous absorption into the event rate, i.e. via the Borrmann effect on the coherency condition of Bragg-Primakoff photoconversion of solar axions. The sensitivity of crystal technologies used in the SuperCDMS, LEGEND, and SABRE setups has been demonstrated, and we find that the inclusion of absorption effects even with Borrmann-enhanced signal rates still would require multi-ton scale detectors to surpass the existing astrophysical constraints in sensitivity to ALPs. However, a dedicated study with a thorough and careful treatment of the absorption suppression and Borrmann effects is definitely needed to better understand its impact on experiments that utilize Bragg-Primakoff conversion. In particular, the evaluation of the imaginary form factors to determine the anomalous absorption effect in materials other than Ge is of particular interest for future work.

Crystal detector technologies are also necessary tools to discriminate axion-like particle signals from other types of BSM and neutrino signatures, with high sensitivity to time modulation from the directional sensitivity of Bragg-Primakoff scattering. This is a powerful tool for background rejection as well, and ideally a joint analysis of multiple detectors situated at different latitudes and longitudes would benefit greatly from leveraging the time modulation of the signal. They are also complimentary to future helioscope experiments like IAXO; while the projected reach for IAXO over the axion-photon coupling parameter space is vast, the sensitivity to solar axions with masses $m_a \gtrsim 1$ eV becomes weaker to coherent Primakoff conversion in magnetic field helioscopes. It was shown in ref.~\cite{Dent:2020jhf} that future liquid noble gas detectors for dark matter direct detection at kilotonne-year scales could begin to probe couplings beyond the astrophysics constraints for axion-like particles, while in this work we find that equivalent reach is possible at tonne-year exposures with crystal detector technology, if utilized to its fullest potential. The presence of complimentary searches at these mass scales is essential for a complete test of the QCD axion parameter space and the parameter space for heavier axion-like particle models.


\let\oldbibitem\bibitem
\renewcommand{\bibitem}{\setlength{\itemsep}{0pt}\oldbibitem}
\bibliographystyle{unsrt}

\phantomsection
\addcontentsline{toc}{chapter}{REFERENCES}

\renewcommand{\bibname}{{\normalsize\rm REFERENCES}}

\bibliography{data/myReference}

%
%
%
%

\begin{appendices}
\titleformat{\chapter}{\centering\normalsize}{APPENDIX \thechapter}{0em}{\vskip .5\baselineskip\centering}
\renewcommand{\appendixname}{APPENDIX}

%
%
%
%


\phantomsection

\chapter{\uppercase{Coherent Atomic Form Factor}}
\label{app:coh_ff}

The atomic form factor, in the original sense, is given by the Fourier transform of the spatial parts of the initial and final state atomic wave functions;
\begin{equation}
    F(q) = \int \psi^*_{\rm f} \psi_{\rm i} e^{i \vec{q}\cdot\vec{r}} d^3 r
\end{equation}
This term can usually factor out of the entire matrix element, as long as there are no spin-dependent pieces in the spatial wave functions, such that the matrix element can be factorized as $(\text{Lorentz structures}) \times (\text{Form Factor})$. The exception to this rule is seen in nucleon scattering which has separate form factors for the electric and magnetic dipole moments, due to the different spin rules for each interaction.

For atomic scattering, typically one considers transitions from the ground state wave function $\psi_0$ to an excited state $\psi_E$, or more simply, in the case that there is no change to the spatial in- and out-states, only a momentum transfer to the whole target. In that case, we can take
\begin{align}
    F(q) &= \int |\psi|^2 e^{i \vec{q}\cdot\vec{r}} d^3 r \nonumber \\
    &\propto \int \rho(\vec{r}) e^{i \vec{q}\cdot\vec{r}} d^3 r
\end{align}
where $\rho$ is the mass or charge distribution that is proportional to the wave function amplitude. There are two form factors used in ref.~\cite{Yamaji:2017pep} which I will discuss later;
\begin{equation}
    f_\gamma (\vec{q}) \equiv \frac{1}{e} \int d^3 \vec{r} \rho(\vec{r})e^{i\vec{q}\cdot\vec{r}}
    \label{eq:ffg}
\end{equation}
\begin{equation}
    f_a (\vec{q}) \equiv k_a^2 \int d^3 \vec{r} \phi(\vec{r}) e^{i\vec{q}\cdot\vec{r}}
    \label{eq:ffa}
\end{equation}
The atomic form factor has the property
\begin{equation}
    f_\gamma (0) = Z
\end{equation}
which normalizes $\rho(r)$ to the total atomic charge, $eZ$. The axion form factor, on the other hand, should be treated carefully. We can derive an analog of the Mott-Bethe relation, which ordinarily relates the X-ray and electron scattering form factors, to relate the X-ray and axion form factors $f_\gamma$ and $f_a$, respectively. I use Bethe's approach of inserting the Laplacian acting on $e^{-\vec{q}\cdot\vec{r}}$ equation and integrating by parts;
\begin{align}
    f_a (\vec{q}) &= k_a^2 \int d^3 \vec{r} \phi(\vec{r}) e^{i\vec{q}\cdot\vec{r}} \\
    &= -\dfrac{1}{q^2} k_a^2 \int d^3 \vec{r} \phi(\vec{r}) \nabla^2 e^{i\vec{q}\cdot\vec{r}} \\
    &=-\dfrac{1}{q^2} k_a^2 \int d^3 \vec{r} \nabla^2 \phi(\vec{r})  e^{i\vec{q}\cdot\vec{r}} \\
    &= \dfrac{ k_a^2}{q^2} \int d^3 \vec{r} \Big[ eZ\delta(r) - \rho(r)] \Big]  e^{i\vec{q}\cdot\vec{r}} \\
    &= \dfrac{e k_a^2}{q^2} \Big[ Z - f_\gamma (\vec{q}) \Big] \label{eq:mottbethe}
\end{align}
where I have used Poisson's equation $\nabla^2 \phi(\vec{r}) = - (eZ\delta(r) - \rho(r))$ in natural units, which includes the nucleus charge density $eZ\delta(r)$ and the electron charge distribution with $\int \rho(\vec{r}) d^3 \vec{r} = eZ$. The final expression in Eq.~\ref{eq:mottbethe} agrees with~\cite{Buchmuller:1989rb}.

Next, it was pointed out by~\cite{Ibers:1958} that while $f_\gamma(0) = Z$, $f_a$ does not approach zero in the $q=0$ limit, as the Yamaji paper suggests. This is because after expanding $f_\gamma(q)$ in powers of q, an even series develops whose leading term is $Z$, cancelling the other $Z$ in the bracket, and subleading term is $\propto q^2$ which gets cancelled by the $q^2$ in the denominator of Eq.~\ref{eq:mottbethe}. See below;
\begin{align}
    f_\gamma (\vec{q}) &= \frac{1}{e} \int d^3 \vec{r} \rho(\vec{r})e^{i\vec{q}\cdot\vec{r}} \\
    &=  \frac{1}{e} \int \rho(\vec{r}) \dfrac{\sin(qr)}{qr} r^2 dr \\
    &= \frac{1}{e} \int \rho(\vec{r}) \Big( 1 - \frac{(qr)^2}{3!} + \frac{(qr)^4}{5!} - \dots  \Big) r^2 dr \\
    &= Z \Big(1 - \frac{q^2}{3!}\braket{r^2} + \frac{q^4}{5!}\braket{r^4} - \dots \Big)
\end{align}
Therefore, inserting this expansion into Eq.~\ref{eq:mottbethe}, we have
\begin{equation}
    f_a (q) = e Z k_a^2 \Big( \frac{1}{3!}\braket{r^2} - \frac{q^2}{5!}\braket{r^4} + \dots \Big)
\end{equation}

%
%
%
%


\phantomsection

\chapter{\uppercase{Photoelectric Primakoff Ionization}}
\label{app:pe_prim}

\begin{figure}[h]
 \centering
      \begin{tikzpicture}
              \begin{feynman}
                 \vertex (o1);
                 \vertex [right=1.4cm of o1] (f1) {\(\gamma\)};
                 \vertex [left=1.4cm of o1] (i1){\(a\)};
                 \vertex [below=1.4cm of o1] (o2);
                 \vertex [below=0.1cm of o2] (o3);
                 \vertex [right=1.4cm of o3] (f2) {\(A^+\)};
                 \vertex [left=1.4cm of o3] (i3) {\(A\)};
                 \vertex [left=1.4cm of o2] (i2);
                 \vertex [above=.5cm of f2] (f3) {\(e^-\)};
        
                 \diagram* {
                   (i1) -- [scalar] (o1) -- [boson] (f1),
                   (o1) -- [boson] (o2),
                   (i3) -- [fermion] (o3),
                   (o3) -- [ fermion] (f2),
                   (i2) -- [fermion] (o2),
                   (o2) -- [fermion] (f3)
                 };
        \end{feynman}
       \end{tikzpicture}
\caption{Tree-level ALP ionization through an inverse Primakoff-like scattering with an atomic system.}
    \label{fig:production}
\end{figure}
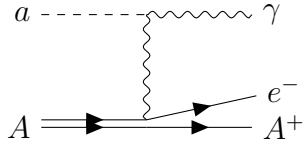

Consider the scattering of an ALP and an atomic system such that the energy transfer of the incoming ALP ionizes the atom, ejecting one of the outer shell electrons into a continuum final state. For the ALP-photon coupling $aF\Tilde{F}$, this can happen through a Primakoff-like process,
\begin{equation}
    a + A \to \gamma + A^+ + e^-
\end{equation}
for initial and final 4-momenta $k_a$, $p_e$, $k_\gamma$, $p_e^\prime$ of the ALP, initial state electron, final state photon, and final state electron, respectively (see also Fig.~\ref{fig:production}).  
Consider Primakoff scattering, but instead of an atomic coherence, we work in the incoherent regime in which a target atom at rest is ionized by the exchange photon. Define the free particle 4-momenta as follows.
\begin{align}
    k_a^\mu \to (E_a, \vec{k}_a) \\ \nonumber
    p_e^\mu \to (E_e, \vec{p}_e) \\ \nonumber
    k_\gamma^\mu \to (E_\gamma, \vec{k}_\gamma) \\ \nonumber
    {p_e^\prime}^\mu \to (E_e^\prime, \vec{p}_e^\prime)
\end{align}
We start with the $2\to 2$ differential cross section element and follow the guidance of refs.~\cite{Catena:2019gfa,Essig:2011nj} for the following calculation.
\begin{align}
    d\sigma =& \dfrac{1}{4 E_a E_e v_a} \dfrac{d^3 \vec{p}^\prime_e}{(2\pi)^3}  \dfrac{1}{2E_e^\prime} \dfrac{d^3 \vec{k}_\gamma}{(2\pi)^3} \dfrac{1}{2E_\gamma} |\mathcal{M}|^2 (2\pi)^4 \delta^4 (k_a + p_e - p_e^\prime - k_\gamma)
 \end{align}
The matrix element can be broken up as the point-wise matrix element combined with the atomic electron matrix element. The electronic part will describe the transition of an electron in the $(n,l)$ shell to a continuum state with momentum $p_e^\prime$ and final state quantum numbers $(n^\prime, l^\prime)$;
\begin{equation}
|\mathcal{M}|^2 = \int \dfrac{d^3 p_e}{(2\pi)^3} \braket{\vec{e}_2 | {\vec{p}_e}^\prime} \braket{|\mathcal{M}_\text{free}(k_a, p_e, k_\gamma, p_e^\prime)|^2} \braket{\vec{p}_e | \vec{e}_1}
\end{equation}
Here, $\braket{|\mathcal{M}|^2}$ is the matrix element for point-wise Primakoff scattering:
\begin{equation}
 \braket{|\mathcal{M}|^2} = \dfrac{e^2 g_{a\gamma}^2}{2 t^2} [m_a^2 t (m_e^2 + s) - m_a^4 m_e^2 - t((s-m_e^2)^2 + st) - t(t-m_a^2)/2],
\end{equation}
Next, we can remove the spatial part of the delta function by absorbing it with the $d^3 p_e^\prime$ integration, making use of the momentum transfer $\vec{q} = \vec{p}_e - \vec{p}_e^\prime$;
\begin{align}
    d\sigma =& \dfrac{V}{4 E_a E_e v_a} (2\pi) \delta (E_f - E_i)  \dfrac{1}{2E_e^\prime} \dfrac{d^3 \vec{k}_\gamma}{(2\pi)^3} \dfrac{1}{2E_\gamma} \int \dfrac{d^3 p_e}{(2\pi)^3} \braket{\vec{e}_2 | \vec{q} + \vec{k}} \braket{|\mathcal{M}_\text{free}|^2} \braket{\vec{p}_e | \vec{e}_1} \\ \nonumber
    =&\dfrac{V \delta(E_f - E_i) }{16 E_a E_\gamma E_e E_e^\prime v_a}  \dfrac{d^3 \vec{k}_\gamma}{(2\pi)^2} \int \dfrac{d^3 p_e}{(2\pi)^3} \braket{\vec{e}_2 | \vec{q} + \vec{k}} \braket{|\mathcal{M}_\text{free}|^2} \braket{\vec{p}_e | \vec{e}_1}
 \end{align}
Now we make a change of variables $d^3 k_\gamma \to d^3 q$ and express $d^3q = 2\pi q^2 dq d(\cos\theta)$, where $\cos\theta$ is the cosine angle between the momentum transfer and the ALP direction. Although the integration over $d^3 p_e^\prime$ was already performed, we need to put back the phase space of the asymptotic free electron final state by acting on $d\sigma$ with the integral operator;
\begin{equation}
    \dfrac{V}{2} \sum_\text{states} \int \dfrac{{p_e^\prime}^3 dT_e}{(2\pi)^3 T_e}
\end{equation}
where $T_e = E_e^\prime - m_e \approx \frac{{p_e^\prime}^2}{2m_e}$. At this stage we can also safely write $E_e E_e^\prime \approx m_e^2$. This gives
\begin{align}
    d\sigma =&\dfrac{\delta(E_f - E_i) }{128 E_a E_\gamma m_e^2  v_a}  \dfrac{2\pi q^2 dq d(\cos\theta)}{(2\pi)^2} \dfrac{dT_e}{T_e} \braket{|\mathcal{M}_\text{free}|^2}|f_{ion}^{nl}(p_e^\prime, q)|^2
\end{align}
taking a familiar definition for the ionization form factor~\cite{Catena:2019gfa,Essig:2011nj};
\begin{equation}
|f_{ion}^{nl}(p_e^\prime, q)|^2 = \dfrac{V4{p_e^\prime}^3}{(2\pi)^3} \sum_{states} \bigg| \int d^3 x \psi_{p_e^\prime, l', m'}^* (\vec{x}) \psi_{n,l,m}(\vec{x}) e^{i \vec{q}\cdot\vec{x}} \bigg|^2
\end{equation}
This ionization form factor has been calculated using a number of schemes, and several codes have been made available by other authors. For this work, we have used [NEEDS CITATION]. We perform the integration over $d(\cos\theta)$ by making use of the relation
\begin{equation}
    \int d(\cos\theta) \delta(E_f - E_i) = \dfrac{E_\gamma}{k_a q}
\end{equation}
after writing the energy delta function in terms of $\cos\theta$ and using the identity $\delta(f(x)) = |f^\prime (x = x_0)|^{-1} \delta(x - x_0)$ where $f(x_0) = 0$. This reduces the cross section element to
\begin{align}
    d\sigma =&\dfrac{E_\gamma}{k_a}\dfrac{1}{128 E_a E_\gamma m_e^2 v_a}  \dfrac{2\pi q dq}{(2\pi)^2} \dfrac{dT_e}{T_e} \braket{|\mathcal{M}_\text{free}|^2}|f_{ion}^{nl}(T_e, q)|^2
\end{align}
The energy prefactors can be written in a manifestly Lorentz-invariant way, since it can be shown that
\begin{equation}
    \dfrac{E_\gamma}{k_a}\dfrac{1}{E_\gamma E_a m_e^2} = \dfrac{4}{(s - m_e^2 - m_a^2)(s - m_e^2 + m_a^2)}
\end{equation}
and, therefore,
\begin{align}
    d\sigma =& [(s - m_e^2 - m_a^2)(s - m_e^2 + m_a^2)]^{-1}\dfrac{1}{64 \pi v_a} q dq \dfrac{dT_e}{T_e} \braket{|\mathcal{M}_\text{free}|^2}|f_{ion}^{nl}(T_e, q)|^2
    \label{eq:dsigma_ion_12}
\end{align}

\begin{figure}[ht]
    \centering
    \includegraphics[width=0.45\textwidth]{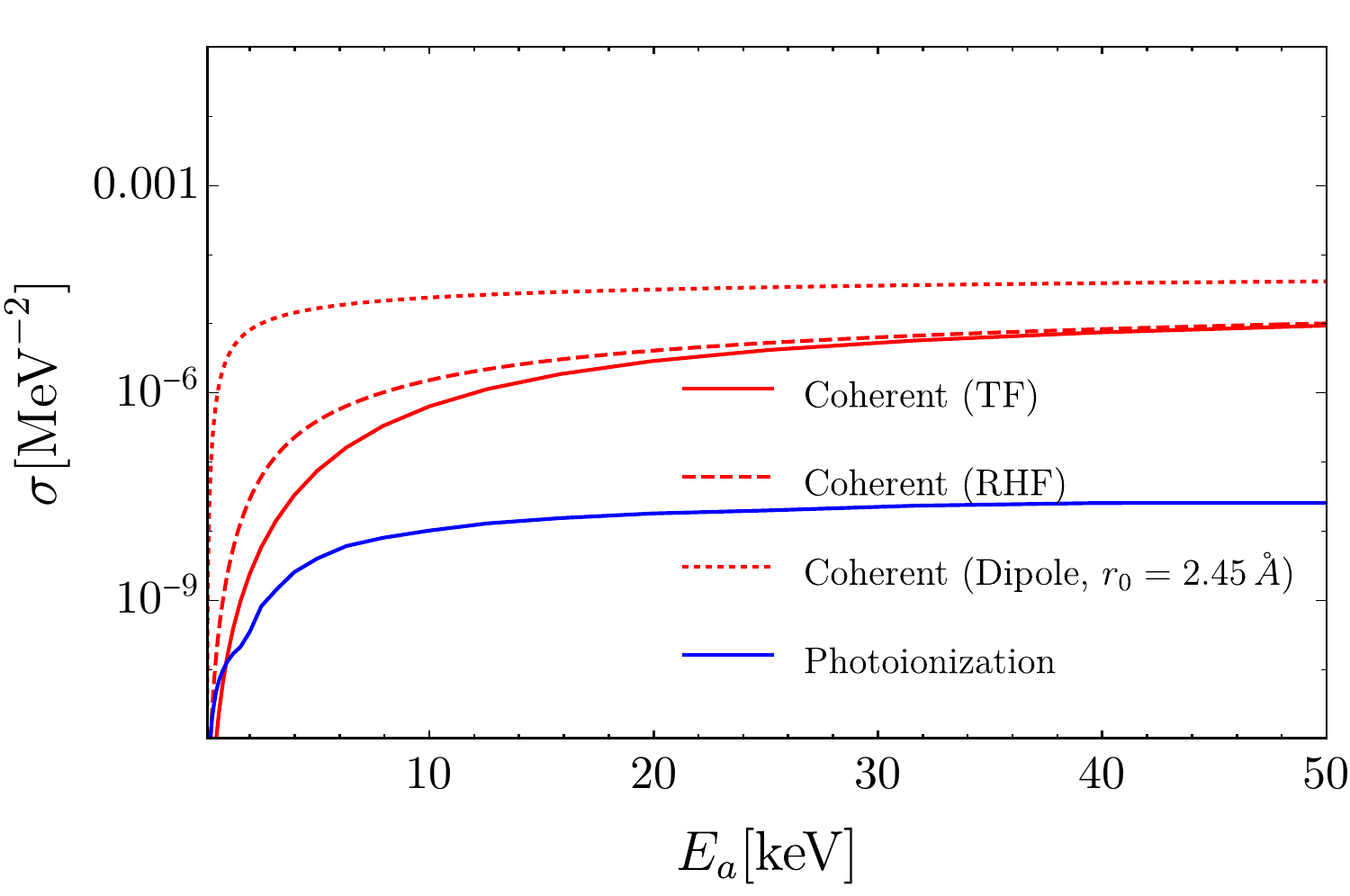}
    \includegraphics[width=0.45\textwidth]{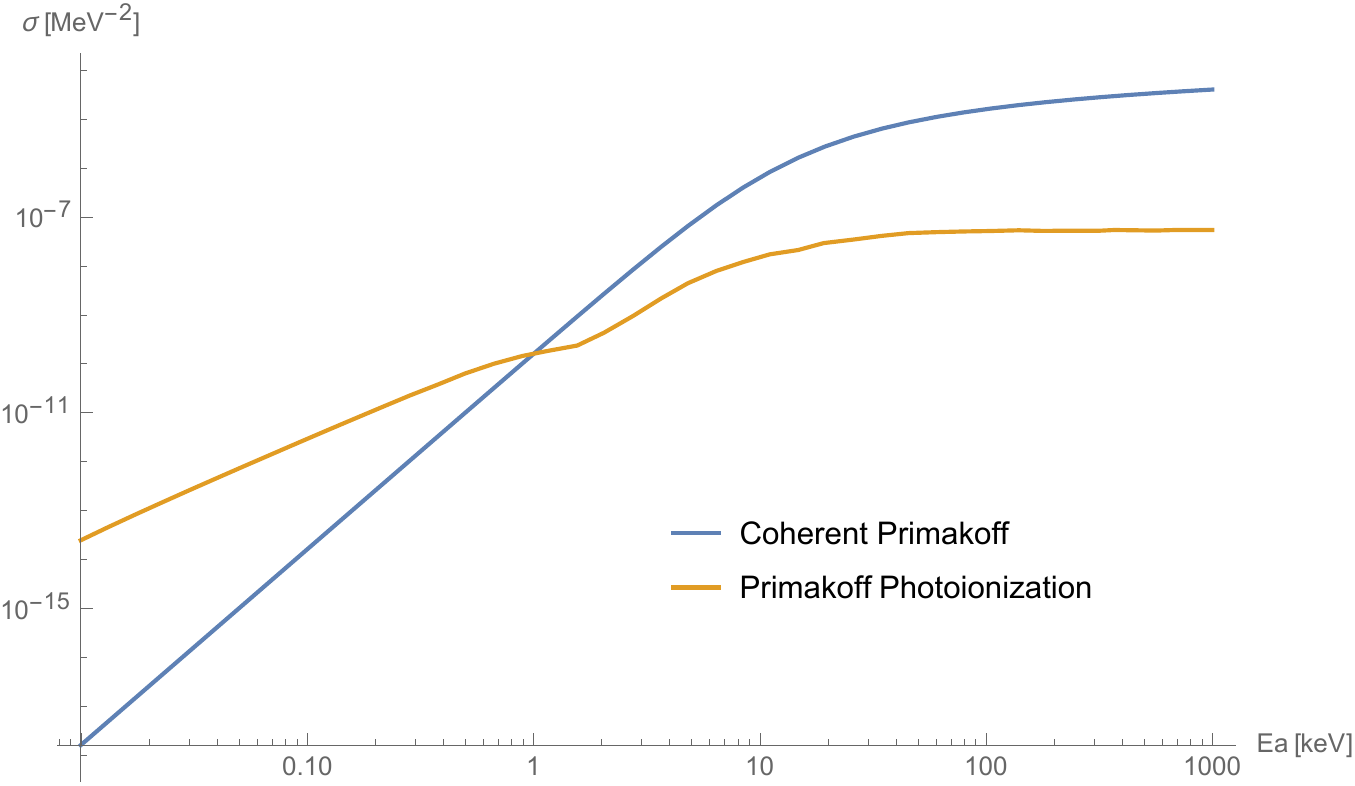}
    \caption{Right: Primakoff scattering cross sections for a variety of atomic form factors (red) compared against the Primakoff photoionization cross section (blue). Left: coherent Primakoff scattering compared with Primakoff photoionization plotted on a log scale to visualize the behavior at low energies where photoionization becomes dominant.}
    \label{fig:primakoff_vs_pe_xs}
\end{figure}

I can connect this expression to the point-wise inverse Primakoff differential scattering cross section, which can be written as
\begin{equation}
    \bigg(\dfrac{d\sigma}{dq^2}\bigg)_\text{point} = \dfrac{1}{16\pi (s-(m_e-m_a)^2)(s-(m_e+m_a)^2)}\braket{|\mathcal{M}_\text{free}|^2} .
\label{eq:dsigma_free}
\end{equation}
Factoring out \ref{eq:dsigma_free} from \ref{eq:dsigma_ion_12} gives us
\begin{align}
    d\sigma =& \dfrac{1}{4 v_a} q dq \dfrac{dT_e}{T_e} \xi(s, m_e, m_a) \bigg(\dfrac{d\sigma}{dq^2}\bigg)_\text{point} |f_{ion}^{nl}(T_e, q)|^2
\end{align}
where we have defined
\begin{equation}
    \xi(s, m_e, m_a) \equiv \dfrac{(s-(m_e-m_a)^2)(s-(m_e+m_a)^2)}{(s - m_e^2 - m_a^2)(s - m_e^2 + m_a^2)}.
\end{equation}
For $m_a^2 << m_e^2$ and $v_a \to 1$, this simplifies to
\begin{align}
    \dfrac{\partial^2\sigma}{\partial T_e \partial q} &=  \dfrac{q}{4T_e} \bigg(\dfrac{d\sigma}{dq^2}\bigg)_\text{point} |f_{ion}^{nl}(T_e, q)|^2 \nonumber \\
    &= \dfrac{1}{8T_e} \bigg(\dfrac{d\sigma}{dq}\bigg)_\text{point} |f_{ion}^{nl}(T_e, q)|^2
\end{align}
This process was investigated further using a different theoretical formalism in ref.~\cite{Wu:2022yei}.

Lastly, let's compare the photoionization process to the coherent Primakoff process. I show this comparison in Fig.~\ref{fig:primakoff_vs_pe_xs}. We see that the usual process without ionization has the dominant cross section in the relativistic / high energy limit, while below 1 keV, ionization begins to dominate. Here I take $m_a \ll 1$ keV.

%
%
%
%


\phantomsection

\chapter{\uppercase {Further ALP Parameter Space Constraints}}
\label{app:fvc}

\begin{figure}[h]
    \centering
    \includegraphics[width=0.8\textwidth]{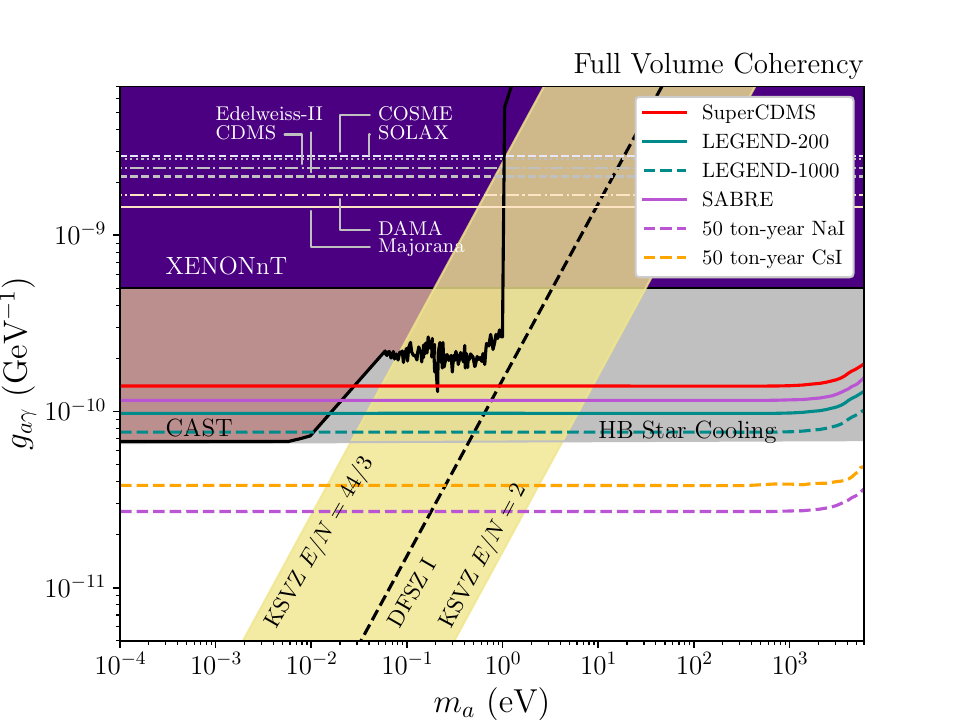}
    \caption{Sensitivity projections for germanium experiments SuperCDMS, LEGEND-200, LEGEND-1000, and SABRE setups with with full volume coherency assumed.}
    \label{fig:sensitivity_with_fvc}
\end{figure}

The existing bounds from DAMA~\cite{Bernabei:2004fi}, CUORE~\cite{Li:2015tyq}, Edelweiss-II~\cite{Armengaud_2013}, SOLAX~\cite{PhysRevLett.81.5068}, COSME~\cite{COSME:2001jci}, CDMS~\cite{CDMS:2009fba}, and Majorana~\cite{Majorana:2022bse} are shown in Fig.~\ref{fig:sensitivity_with_fvc} assuming full volume coherence (FVC), as well as the projected limits for SuperCDMS, LEGEND-200, LEGEND-1000, SABRE, and future CsI and NaI detectors with FVC. However, the assumption of FVC in previous experiments and the forecasted limits shown is likely poor, and depending on the module size and material assumed in each case, the attenuation factor $I(\vec{k},\vec{G})$ will give varying amounts of suppression to the event rate, changing the sensitivites and existing exclusions shown here.

\begin{figure}[h]
    \centering
    \includegraphics[width=0.8\textwidth]{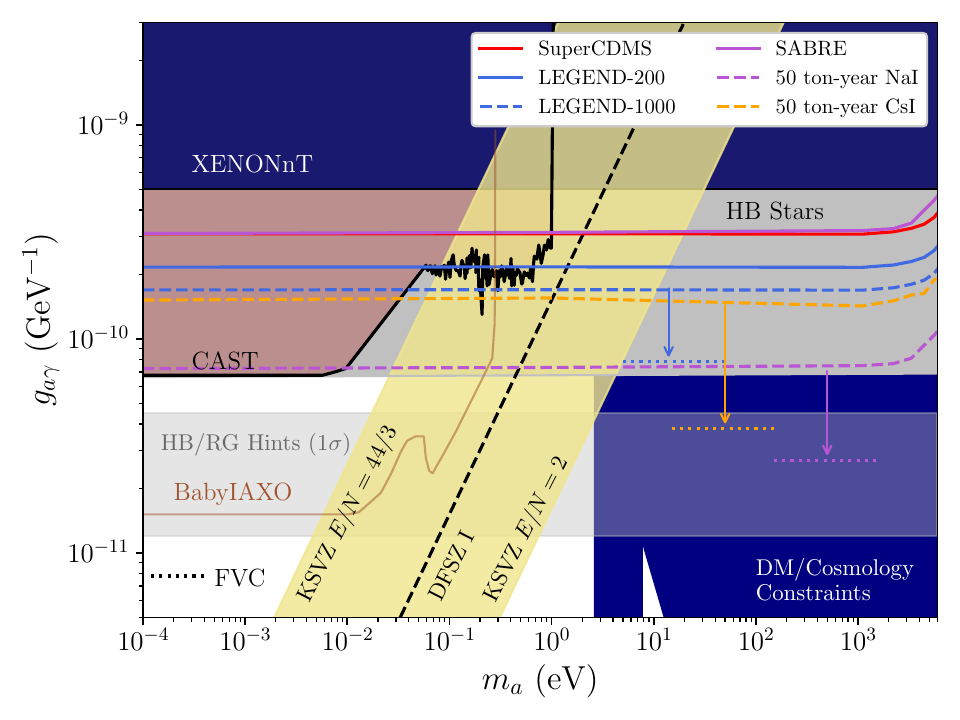}
    \caption{Sensitivity projections for germanium experiments SuperCDMS, LEGEND-200, LEGEND-1000, and SABRE setups with with the stellar cooling hints and constraints on DM axions and from cosmology.}
    \label{fig:sensitivity_with_hints_with_dm}
\end{figure}

Next, in Fig.~\ref{fig:sensitivity_with_hints_with_dm} I show the projections with aborption effects and with FVC (arrows). I additionally overlay the red giant and horizontal branch cooling hints (at $1\sigma$ for a pure $g_{a\gamma}$ coupling) and the constraints from assuming ALPs make up the dark matter and constraints from cosmological considerations.

\end{appendices}

\end{document}